\documentclass[12pt,preprint]{aastex}
\usepackage{subfigure}
\usepackage{rotating}
\newif\ifAMStwofonts
\AMStwofontstrue

\def\ga{\mathrel{\hbox{\rlap{\hbox{\lower4pt\hbox{$\sim$}}}\hbox{$>$}}}}
\def\la{\mathrel{\hbox{\rlap{\hbox{\lower4pt\hbox{$\sim$}}}\hbox{$<$}}}}

\shorttitle{Pulsar Precession and ISM Magnetic Field Variations}

\shortauthors{J. M. Weisberg  et al.}

\begin{document}

\title{A Search for Neutron Star Precession and Interstellar Magnetic Field Variations via 
Multiepoch Pulsar Polarimetry}

\author{J. M. Weisberg$^{1}$, J. E. Everett$^{1,2}$, J. M. Cordes$^{3}$, J. J.  Morgan$^{1}$, 
and D.  G. Brisbin$^{1,3}$}
\setcounter{footnote}{1}
\footnotetext{Department of Physics and Astronomy, Carleton College,
Northfield, MN 55057}
\setcounter{footnote}{2}
\footnotetext{Departments of Physics and Astronomy, and Center for
 Magnetic Self-Organization in Laboratory and Astrophysical Plasmas,
 University of Wisconsin, Madison, WI 53706}
\setcounter{footnote}{3}
\footnotetext{Department of Astronomy and National Astronomy \& Ionosphere 
Center, Cornell University, Ithaca, NY 14853}

\begin{abstract}

In order to study precession and interstellar magnetic field
variations, we measured the polarized position angle of 81 pulsars at
several-month intervals for four years. We show that the uncertainties
in a single-epoch measurement of position angle is usually dominated
by random pulse-to-pulse jitter of the polarized subpulses. 
Even with these uncertainties, we find that the position angle
variations in 19 pulsars are significantly better fitted (at the
3$\sigma$ level) by a sinusoid than by a constant.  Such variations
could be caused by precession, which would then indicate periods of
$\sim(200-1300)$~d and amplitudes of $\sim(1-12)$~degrees.  We narrow
this collection to four pulsars that show the most convincing evidence
of sinusoidal variation in position angle.Also, in a handful of
pulsars, single discrepant position angle measurements are observed
which may result from the line of sight passing across a discrete
ionized, magnetized structure.  We calculate the standard deviation of
position angle measurements from the mean for each pulsar, and relate
these to limits on precession and interstellar magnetic field
variations.

\end{abstract}


\slugcomment{Accepted by APJ 15 Jul 2010}

\setcounter{footnote}{0}

\section{Introduction}

Many pulsars are among the most highly polarized astrophysical sources of radiation.  One polarized
parameter, the position angle of linear polarization $\psi$ (also called the PPA -- polarized position 
angle) is a particularly useful quantity:  it carries information about the intrinsic geometry of the
emitting region at the pulsar [e.g., via the rotating vector model \citep{rc69}]; it also
probes the magnetized
plasma along the line of sight through Faraday rotation measurements [e.g., \citet{7}]. The 
purpose of this paper is to study the behavior of the polarized position angle
on timescales of $10^{2-3}$ d, in order to investigate  time-variable behaviors of pulsars' 
magnetospheric geometry and the interstellar medium.
For example, precession of the neutron star or other long-timescale oscillatory behavior 
affecting the emission beam
will change the PPA, as will variations in the magnetized interstellar plasma along the line 
of sight.  

\section{Observations and Analyses}

\label{sec:obsana}
Ninety-eight pulsars were observed at 21 cm from Arecibo Observatory every few months for several  
years from 1989 to 1993.  The resulting grand average polarized pulse profiles were analyzed in 
\citet{w99}, where observing details can also be found.  In the present work, we found that we had 
adequate data on eighty-one of the pulsars to search for  temporal variations  of polarized position
angle.  For all of them, the polarized position 
angles were originally measured relative to an unknown origin.  Consequently,
in the present work,  in order to search for temporal  {\it{changes}} in position angle, we needed
to reference our data to a position-angle  standard.  To do so, we selected two calibrator pulsars,
B0540+23 (J0543+2329) and B1929+10 (J1932+1059).   The calibrators were chosen on the 
basis of the relative constancy of their
PPA's during full-sky Arecibo observing tracks.  We then  compared the measured polarized 
position angle of a 
target pulsar on a given observing session to the PPA of one of the  two calibrator pulsars.  
While the
{\it{absolute}} position angle is still unknown, 
this referencing procedure enables us to be 
sensitive to  position angle {\it{changes}} over time, assuming only 
that the position angle of the calibrators remains fixed over time.   The
procedures are described in the following sections; mathematical details are given in Appendix A.

In the rare cases where a calibrator pulsar was not observed during a daily session, we created
an artificial ``meshed''  calibrator pulsar by determining its expected position angle on that session 
from observations of all other pulsars during that session.

\subsection{A Pulsar's Polarized Position Angle Curve}

Pulsars tend to be very highly linearly polarized objects. 
The observed linear polarization originates at or near the star, often following the magnetic
field lines at the point of origin \citep{rc69}, and is then modified by various propagation effects.     
We define the generic  polarized position angle ``curve "  for a pulsar at epoch $t$, a function of 
pulsar longitude $\phi$ at fixed $t$, as the set of polarized
position angles $\psi(\phi,t)$ for $0 \le \phi \le 2 \pi$.  (See  the bottom panels of 
Fig. \ref{fig:obsPAs}  for  examples of PPA curves.)

The measured PPA curve 
at epoch $t$ can be expressed as the sum of
five terms -- an ensemble
average (denoted below by angular brackets) for the indicated
pulsar longitude $\phi$ and epoch $t$ which reflects the pulsar's intrinsic orientation, 
an interstellar Faraday rotation term, an instrumental origin,
and a term arising from  errors  due to radiometer noise and one from
intrinsic pulse-to-pulse 
fluctuations in amplitude and phase of individual pulses
(called ``jitter;'' see \citet{c93} and further discussion below):
\newcommand{\psibarp}{\overline\psi_p}

\begin{equation}
\psi(\phi,t)=
\langle \psi(\phi, t) \rangle +
	\psi_{Faraday}(t)+
	 \psi_{instr}(t) +
	\delta\psi_{noise}(\phi, t) +
	\delta\psi_{jitter}(\phi, t).
\label{eqn:basicpsi}
\end{equation}
The first two terms on the right-hand side of the above equation hold the physics of interest 
in this experiment -- namely the temporal behavior of the pulsar's orientation or magnetosphere,
and of  the interstellar magnetic field through which its signals propagate, respectively.  The
third term defines the  instrumentally
measured  PPA corresponding to an absolute PPA on the sky of zero. The latter two terms, 
though having zero mean,  represent phenomena  that add variance to the 
measurements, thereby limiting our precision in specifying the first two. We now discuss each
of these five terms in order.

\subsubsection{Variations in PPA Due to Changes in Orientation of the Pulsar or its Magnetosphere}
\label{sec:orientation}
{\bf{(a) Precession:}}  Precession would lead to a periodic wobble of the pulsar orientation.  
Most pulsar magnetospheric
models indicate that the polarized emission is tied fundamentally to the geometry of the magnetic
field in the emitting
regions. Hence it is expected that a periodic wobble in pulsar orientation translates directly into a 
periodic variation of similar amplitude in polarized position angle.  
Therefore, a search for
temporal variations in PPA is  a sensitive probe for the presence of precession. 

\begin{figure}[p]
\begin{center}
\subfigure[]
{     \includegraphics[height=7cm, trim= 2in 2in 2in 2in ]{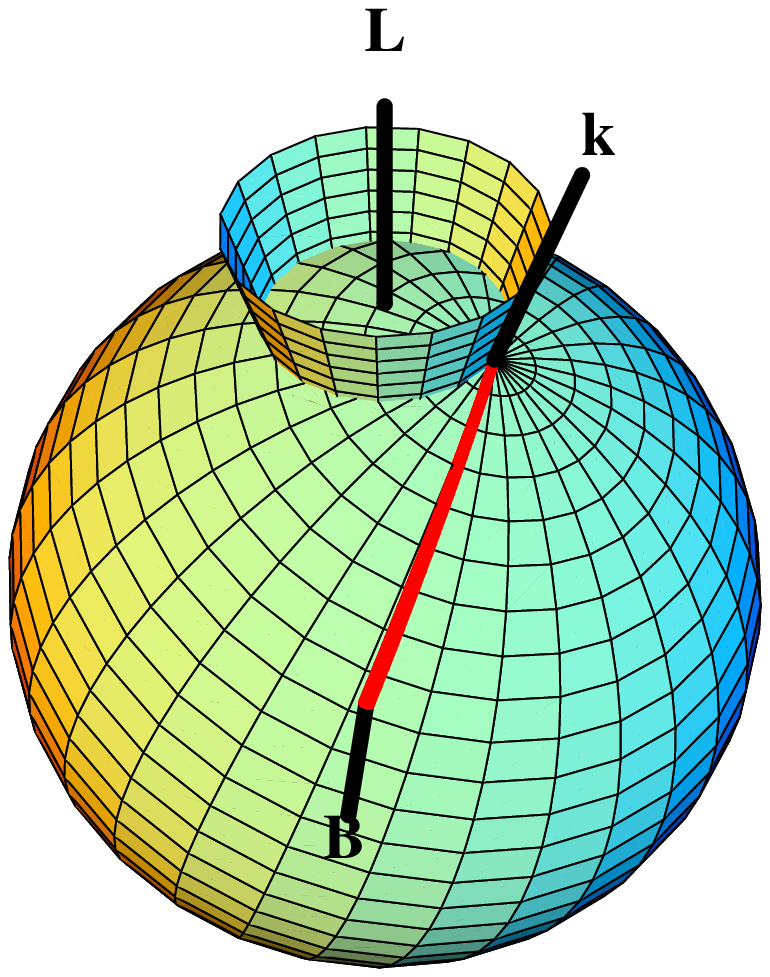}      }
\qquad
\subfigure[] 
{     \includegraphics[height=7cm, trim= 2in 2in 2in 2in  ]{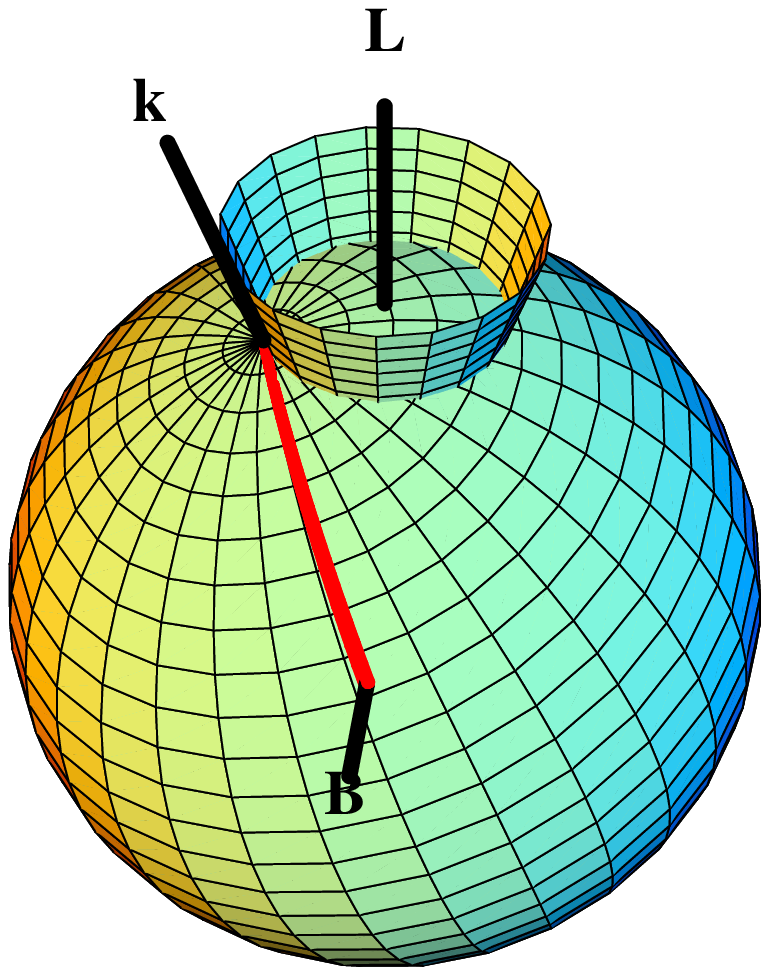}   }
\caption{Two panels from the animation starPrecession.mpg, which is available at 
\underline{this link}.  Each panel of the animation shows a precessing 
pulsar stroboscopically at  times  (separated by multiples of 
$\sim$ 1 rotational period)  that its beam  points most directly at the observer; i.e., at those 
times when it is observable.   The total angular momentum vector, $\bf{L}$, is conserved, while the 
body symmetry axis $\bf{k}$  describes a cone about $\bf{L}$ approximately once per spin 
period. The absence of perfect synchronization between these two rates leads to a slow 
apparent precession of $\bf{k}$   at successive observer viewing times.  The above two 
panels from the animation were chosen to illustrate the pulsar geometry at  the extrema of its 
precessional excursion.    The beam / magnetic axis is $\bf{B}$. Note 
the changing orientation of the red  meridian connecting $\bf{k}$ and $\bf{B}$ on precessional
timescales.  This meridian 
represents  a fiducial magnetic field line projected onto the surface. In most pulsar models, linearly
polarized emission will be oriented along (or perpendicular to) this meridian, so the precessional 
motion leads  to a changing polarized position angle (PPA) on precessional timescales.}
\label{fig:animation}  
\end{center}  
\end{figure}

The simplest model is one where the wobble would result from the free precession
of an asymmetric isolated pulsar, although there also exist models where the
precession is forced by an external disk or companion \citep{qxxwx03}, or by the star's own 
radiation \citep{m00}. We assume that
the neutron star is approximately spherical, with a slight axially symmetric oblateness about a 
symmetry axis.  Hence the three moments of inertia have the relationship 
$I_1\equiv I_{||} \gtrsim I_2=I_3$, where $I_{||}$ is the moment parallel to the symmetry axis. (For
extension of the precession model to a triaxial body, see \citet{alw06}).
\citet{s86} and \citet{ja01} show that the motion of a point on the star can be decomposed into a 
quick rotation
about the total angular momentum vector plus a slow precession about the body symmetry axis.
For a sketch of the process and a link to an animation, see Fig. \ref{fig:animation}; for readers wishing to 
modify the geometrical
parameters of the animation, the Mathematica source code is also
available online at  the Mathsource library \citep{ew07}.  The angle $\theta$ between the angular 
momentum vector and symmetry axis leads to a ``wobble'' at both the rotation
and precession frequencies, with the latter leading directly to a  change in PPA of similar 
amplitude\footnote{The amplitude of the PPA variation, $ \Delta\psi$, is magnified by a factor of 
$\sim [\cos \alpha]^{-1}$; .i.e, $\Delta\psi\sim \theta \times[\cos \alpha]^{-1}$,
where $\alpha$ is the (generally unknown) latitude of the magnetic axis with respect to the  
symmetry equator.} on precession timescales.  Very slow precession, with a period of many years,
would produce just a linear trend in PPA with epoch.

{\bf{(b)Magnetospheric Adjustments:}} As we will describe below, there are theoretical difficulties 
(though not insoluble ones) with the picture of 
precession-induced pulse shape variations, centered primarily on the long observed timescales 
and relatively
large inferred amplitudes.  \citet{rg06} have proposed an alternate model whereby the 
pulse shape and timing variations
are caused by  slowly rotating magnetospheric currents rather than by precession of the whole
star.  While this is an interesting alternative interpretation of the observations, more work needs 
to be done to flesh out the model and to verify that it can explain the observed timescales and 
amplitudes.

\subsubsection{Variations in PPA Due to Changes in Interstellar Faraday Rotation}

\label{sec:faraday}

In this section, we focus on studying variations in the magnetized ISM plasma via 
a search for temporal variations in PPA.  The second term of Eq. \ref{eqn:basicpsi} quantifies 
Faraday rotation of the PPA in the interstellar medium, $\psi_{Faraday}(t)$, at some epoch, $t$.  
The rotation measure,
$RM(t)$, is defined so as to specify the angle of PPA  Faraday rotation at wavelength 
$\lambda$:
\begin{equation}
\psi_{Faraday}(t) = \lambda^2 RM(t),
\label{eqn:faradayrot}
\end{equation}
with angle in radians, $\lambda$ in m, and $RM(t)$ in rad m$^{-2}$.

A measured change in  PPA, $\Delta \psi_{Faraday}$, then implies a 
change in Faraday rotation measure, $ \Delta RM$:
\begin{equation}
 \Delta RM=\frac{\Delta\psi_{Faraday}  }{  \lambda^2 }.
\label{eqn:Deltafaradayrot}
\end{equation}

In order to relate $\Delta RM$ to the ISM, 
we express $RM$ as a function of fundamental properties 
along the pulsar--Earth path:  
\begin{equation}
RM = 0.81 \int_0^{D} B_{||}(s)\  n_e(s)  \ ds,
\label{eqn:faradayB}
\end{equation}
where 
 $B_{||}$ is the magnetic field parallel to the line of sight in $\mu$G, $n_e$ 
 is the electron density in electrons\,cm$^{-3}$, and $ds$ is a differential path element  
 whose integrated distance is $D$ \citep{lgs06}.

Each of the variables in Eq. \ref{eqn:faradayB} may be a function of time.
While this time-variability could result directly from {\it{temporal}} changes
in the ISM, it would more likely be induced  by
the pulsar's motion rapidly sweeping the line of sight across a
presumed {\it{spatially}} inhomogeneous ISM. 

Since  measurements show that  temporal variations of 
dispersion measure  $ DM  =  \int_{0}^{D}{n_e ds}$, are generally quite small
\citep{pw91,bET93,hET04,7,aET05,yET07}, we further expect that any 
observed variations in $RM$ would originate principally from   the pulsar carrying
the line of sight  across an inhomogeneity in $B_{||}$.
Therefore, we model the RM change as resulting from the line 
of sight passing across a discrete region
having a variation in the parallel field, $\Delta B_{||}$, occupying a fraction 
$f$ of the pulsar-Earth path, with {\it{no}} associated $n_e$ variation.

In this case,  $\Delta RM$, the $RM$ change due to the discrete region is
\begin{equation}
\Delta RM = 0.81  \int_0^{f D} \Delta  B_{||} \    n_e  ds 
\approx                              0.81 \  \Delta B_{||} \  f \ DM.
\label{eqn:DeltaRMfromgrad}
\end{equation}

Finally, we can relate an observed change in PPA, $\Delta\psi_{Faraday}$, 
to the properties of the discrete region by combining
Eqs. \ref{eqn:Deltafaradayrot} and \ref{eqn:DeltaRMfromgrad}:
\begin{equation}
f \ \Delta  B_{||}  \approx \frac{\Delta RM}{0.81 \ DM}
=   \frac {\Delta\psi_{Faraday}  }{   0.81 \  \lambda^2  \  DM};
\end{equation}
or, expressed as a fractional variation:
\begin{equation}
\frac{f \ \Delta B_{||}   }{  \langle B_{||} \rangle  } \approx
\frac{\Delta RM  }{  RM} = 
\frac{  \Delta\psi_{Faraday}         }{ \lambda^2   RM},
\label{eqn:deltaRMfrac}
\end{equation}
where  $\langle  B_{||}\rangle$ is the line of sight average.

\subsubsection{The Instrumental PPA and its Variance}

The third term in Eq. \ref{eqn:basicpsi} represents a conversion of absolute PPA on the sky, to 
PPA measured in the instrumental
frame.  The instrumental frame involves geometrical terms associated 
with telescope orientation as well as instrumental phase delays during  a given session.   As 
noted above in \S\ref{sec:obsana}, we did not directly determine the instrumental PPA but rather 
referenced our  measured PPA to the PPA of a calibrator pulsar, which has its
own variance (see \S\ref{sec:offset}).

\subsubsection{The Variance in PPA due to Radiometer Noise}

The noise term  in Eq. \ref{eqn:basicpsi} results from the finite sampling statistics of a noisy 
signal and is familiar to
all radioastronomers.  Its contribution to the variance, $\sigma^2_{\psi, \rm\ noise}$, 
was determined empirically by measuring the variance off-pulse.

\subsubsection{The Variance in PPA due to Pulse Jitter}

The jitter term  in Eq. \ref{eqn:basicpsi} is unique to pulsars and requires further explanation,
especially because it often dominates the  error budget.  We characterize as ``jitter'' all 
PPA changes on pulse-to-pulse timescales that are counterparts to the arrival-time jitter
of individual subpulses. These random PPA variations occur due to various propagation effects
in the pulsar magnetosphere  analogous to Faraday rotation. These fluctuations
might be correlated, for example, for the duration  of a single  subpulse 
in a given pulse period, but would be uncorrelated {\em{between}} subpulses
and  {\em{across}} pulse periods.    Random variations in {\em altitude} of emission would 
produce a PPA curve that is shifted earlier or later  for 
higher or lower altitudes, respectively.   This kind of
variation would also be highly correlated across a single subpulse. The variance of $\psi$
due to jitter, $\sigma^2_{\psi, \rm\ jitter}$, is calculated as follows.

\label{sec:jittervar}
By definition, uncertainties  $\sigma_{I, \rm\ jitter}(\phi)$ and 
$\sigma_{L, \rm\ jitter}(\phi)$ in polarized parameters 
$I(\phi)$ (total power at longitude bin $\phi$) and $L$ (linearly polarized power at 
longitude bin $\phi$)  due to random pulse-to-pulse  jitter are 
characterized by
\begin{equation}
\sigma_{I, \rm\ jitter}(\phi)  \equiv m_I(\phi) I (\phi),
\end{equation}
and
\begin{equation}
\sigma_{L, \rm\ jitter}(\phi) \equiv m_L(\phi) L(\phi),
\end{equation}
where $m_X$ is the ``modulation index" of  polarized parameter $X$, a measure of 
its pulse-to-pulse modulation.

In our experiment, each polarized profile  was integrated online for two minutes, thereby
attenuating these pulse-to-pulse modulations.  Therefore,  the jitter-induced fluctuations 
in the measurement of these quantities after $N$ pulses, will be 

\begin{equation}
\sigma_{I, \rm\ jitter}(\phi,N ) =  \sigma_{I, \rm\ jitter}(\phi)/ \sqrt{N} = m_I(\phi) I (\phi) / \sqrt{N},
\label{eqn:sigma_I}
\end{equation}
and
\begin{equation}
\sigma_{L, \rm\ jitter}(\phi,N) =\sigma_{L, \rm\ jitter}(\phi)  / \sqrt{N} = m_L(\phi) L(\phi)  / \sqrt{N}.
\label{eqn:sigma_L}
\end{equation}

At high $S/N$ and under the reasonable assumption that $\sigma_Q\approx \sigma_U$,  
$\sigma_L$ propagates to an uncertainty in position angle $\sigma_{\psi}$:
\begin{equation}
\sigma_{\psi}(\phi)  \approx \sigma_L(\phi)  / (2 L(\phi) )
\label{eqn:sigmapsiphi}
\end{equation}
 \citep{nkc93}.
Therefore, the jitter-induced uncertainty  in the measurement of PPA in an integration
of $N$   pulses is
\begin{equation}
\sigma_{\psi,  \rm jitter}(\phi,N)  \approx m_L(\phi) / (2 \sqrt{N})
\end{equation}

As we shall see below in \S \ref{sec:offset}, we must ultimately
determine $\sigma_{\psi, \rm jitter}$, the jitter-induced fluctuation
in the PPA integrated not only over $N$ pulses but also over the full
range of longitudes.  With the subpulse as the basic unit of
correlated emission, the jitter-induced fluctuations in the
measurement of PPA after $N$ pulses, each containing $n$ subpulses,
will be characterized by
\begin{equation}
\sigma_{\psi,  \rm jitter}=\sigma_{\psi,  \rm jitter}(N,n)=\sigma_{\psi,  \rm jitter}(\phi,N)/\sqrt{n}  
\approx m_L / (2 \sqrt{N n}).
\label{eqn:sigmapsijitt}
\end{equation}

Since we could not directly measure the quantities $m_L$ and $n$ with
our two-minute integrations, we surveyed the literature for typical
values.  Many observations suggest that $m_I\sim 1$
\citep{bET80,wET86,wET06} and the few extant polarized modulation
index observations (e.g., \citet{j01}) support the reasonable
assumption that $m_L\sim 1$ as well.  Single-pulse observations
indicate that $n$ is also approximately 1.  Finally, we found that
constant-$\psi$ fits to the full set of pulsars had a mean
$\chi_{\nu}^2 \sim1$ when $m_L/\sqrt{n}\sim 1.7$ in
Eq. \ref{eqn:sigmapsijitt}, so we adopted this value; i.e.,

\begin{equation}
\sigma_{\psi, jitter}\sim 0.85/\sqrt{N}.
\label{eqn:jitterempirical}
\end{equation}

We empirically tested the relative contributions  of radiometer noise and pulse jitter to the PPA
uncertainty in the following tests on one of the two calibrator pulsars, B0540+23. We 
observed it for 
several  $\sim$two-hour stretches with a series of standard two-minute
integrations identical in form to all others in this experiment, and analyzed the resulting
data as follows.  First, a linear trend in PPA, 
presumably due to variable ionospheric Faraday rotation,  was fitted and subtracted out
\footnote{To first approximation, ionospheric Faraday rotation is nulled in our usual pulsar -- 
calibrator pair measurements, but not in these all--sky measurements of a single pulsar.}.
For PSR B0540+23, the uncertainty in PPA  resulting from  radiometer noise was $
\sigma_{\psi, \rm noise}\sim0\fdg2$, while the {\it{observed}} standard deviation of two-minute 
integrations from the linear trend was  a far larger 
$\sim1\degr$. Pulse jitter considerations, however, lead via Eq. \ref{eqn:jitterempirical} to an 
estimated $\sigma_{\psi, \rm jitter}\sim2\degr$.  

While the jitter uncertainty estimated via
Eq. \ref{eqn:jitterempirical} appears to be about twice the observed
standard deviation, it is important to bear in mind that the
simplifying assumptions used in the jitter calculation can easily lead
to factor of $\sim2$ errors in either direction for a particular
pulsar, as they apparently have in this case. (See further discussion
in \S\ref{sec:upperlimits}).  Nevertheless, the important conclusion
of this exercise is that the actual PPA uncertainty is far larger than
that due to radiometer noise alone, and that pulse jitter appears to
be the predominant source of the observed noise, with an amplitude in
rough agreement with our model. Note also that the jitter model should
give excellent results for a given pulsar if all factors in
Eq. \ref{eqn:sigmapsijitt} are directly measured, rather than using an
average over our pulsar sample, as in Eq. \ref{eqn:jitterempirical}.

\subsection{Estimation of a Pulsar's Polarized Position Angle Offset with Respect to a Calibrator Pulsar, 
and its Variance}

\label{sec:offset}
In order to study temporal variations in a pulsar's orientation or in the intervening medium, we 
investigate the PPA as a function of time.  As noted in \S\ref{sec:obsana}
, the instrumental PPA origin, 
$\psi_{instr}(t)$ of Eq. \ref{eqn:basicpsi}, was not explicitly
measured.  Instead, we reference our target pulsar PPA's to  ``calibrator'' pulsars B0540+23
or B1929+10, which can be considered PPA ``beacons" of constant but unknown PPA.  We do
so in a three-step process:  first we find the offset of the target pulsar's PPA curve at epoch $t$ 
from a high $S/N$
``template'' PPA curve of the same pulsar from \citet{w99} (see Fig. \ref{fig:obsPAs}a and Eq. 
\ref{eqn:Deltapsipsr}).  Next we do the same procedure at the same epoch on the calibrator 
pulsar B0540 or B1929+10  (see Fig. \ref{fig:obsPAs}b and 
Eq. \ref{eqn:Deltapsical}).
Finally, to yield the desired offset of the target pulsar from the calibrator pulsar at epoch $t$, 
$\Delta\psi(t)$, we difference the above two results (see Eq. \ref{eqn:Deltapsiwhole}).

The variance in a target pulsar-template PPA difference measurement,  
$\sigma^2_{\psi (t) }$, is  the 
sum of the noise and jitter contributions from the target pulsar at epoch  $t$ 
(the template contributes negligibly):  
\begin{equation}
\sigma_{\psi(t)}^2=
\sigma_{\psi(t),\rm\ noise}^2 + \sigma_{\psi(t),\rm\ jitter}^2.
\label{eqn:sigmaDeltapsi}
\end{equation}
A similar equation holds for the variance in a calibrator pulsar-template PPA difference 
measurement,
$\sigma^2_{\psi_{\rm{cal}}(t)  }$.  However, we are ultimately interested 
in the variance of the position angle of a pulsar with respect to the calibrator pulsar, 
({\it{cf.}} Eq.  \ref{eqn:Deltapsiwhole}). This variance, $\sigma^2_{\Delta\psi(t)  }$,  will 
then be the  sum of the above-discussed variances in the pulsar-template and 
calibrator-template PPA difference measurements:
\begin{equation}
\sigma_{\Delta\psi(t)  }^2=
 \sigma_{\psi(t)}^2 + \sigma_{\psi_{cal}(t)}^2.
\label{eqn:sigmapsiml}
\end{equation}

\section{Results}

We searched for evidence of temporal variations of polarized position angle in each of the 81 
target pulsars, with an emphasis on a few likely signatures:  (1)  sinusoidal variations in PPA
due to periodic phenomena at the pulsar such as precession, as described in 
\S\ref{sec:orientation};  (2) discontinuous changes in PPA
caused by interstellar propagation effects, such as the line of sight crossing a  region 
with significantly different magnetic field, as discussed in \S\ref{sec:faraday};
and (3) ramps in PPA caused by phenomena such as those listed above, having a characteristic
time longer than our experiment's.  For each  pulsar, the dataset consists of
the measured PPA offsets, $\Delta\psi(t)$ ({\it{cf.}} Eq. \ref{eqn:Deltapsiwhole}) , and their 
estimated uncertainties,   $\sigma_{\Delta\psi(t)}  $ ({\it{cf.}} Eq. \ref{eqn:sigmapsiml}), at a 
set of epochs, $t$.  In the following sections, we give an overview of these results.  See the
Discussion (\S\ref{sec:discussion})  for exploration of the astrophysical implications.

\subsection{Sinusoidal Variations of Polarized Position Angle}

\label{sec:sine}
To test for possible periodic variations in each pulsar's PPA, we 
fit a sine function to
$\Delta \psi(t)$.   This sine function requires four free parameters:
the zero-point $\Delta \psi_{\rm0}$ about which the sinusoid oscillates; 
the amplitude $a_{\psi}$ of the sinusoid about $\Delta \psi_{\rm
0}$; the period $P_{\psi}$ of the sinusoid; and a phase offset $t_0$:
\begin{equation}
\Delta \psi(t) = \Delta\psi_0 + a_{\psi}\sin [2\pi(t-t_0)/P_{\psi}]. \label{eq:sineEq}
\end{equation}
In fitting with these free parameters, we had two chief concerns: that
our best-fit results for this function's parameters not be dependent
on initial guesses for those parameters, and that we only accept the
sinusoidal fit's results when that fit represents a statistically
significant improvement over the fit of a constant, $\overline{\Delta
\psi}$.

To ensure that the fit results were not dependent on initial guesses,
we ran trial fits on a large number of initial values for both
sinusoidal period $P_{\psi}$ and amplitude $a_{\psi}$.  The
initial-parameter search space for the amplitude was set to lie in the
interval [$0^\circ, 2\cdot \rm{max} (\Delta\psi(t) -
\overline{\Delta\psi})$].  Meanwhile, the maximum sinusoidal period
was set to $5 \cdot \Delta t$, where $\Delta t$ is the total timespan
of the observations, in order to look for long-timescale variations.
The minimum sinusoidal period tested was to $2\cdot \Delta t/N_{\rm
sessions}$ where $N_{\rm sessions}$ is the number of observation
sessions for each pulsar.  This range of sinusoidal amplitude and
period was then used to define a linearly-spaced grid of
$501\times501$ initial guesses.  A separate non-linear least-squares
fit was run for each set of initial parameters, where all four
parameters were allowed to ``float'' freely; the only exception to
this was that the period was not allowed to fall below the minimum
period already defined.  Of all of the resultant converged fits, the
sinusoid with the minimum $\chi^2$ was selected as the best fit for
that pulsar.  As a final check, we re-ran the analysis with a
$4001\times4001$ grid of initial guesses; no significant changes to
our fit results were found.

To check that the best sinusoidal model significantly improves the fit
when compared to the constant-PPA model, we applied the F-Test
\citep[see \S 11.4,][]{br92}.  For the best-fit sinusoid for each
pulsar, we calculated the $F_{\chi}$ statistic, where $F_{\chi} =
\Delta \chi^2/\chi^2_{\nu}$. Here, $\Delta \chi^2$ is the difference
in $\chi^2$ between the best-fit constant PPA and the best-fit
sinusoid, and $\chi^2_{\nu}$ is the reduced $\chi^2$ value for the
best-fit sinusoid.  To assess its significance, one calculates the
probability that the sinusoidal model's improved fit would be achieved
with random data.  To accept it as formally significant, we ask that
the sinusoidal fit be an improvement over the constant-PPA fit at the
$P\sim99.7\%$ level (approximately $3\sigma$); i.e., we require that
the calculated improvement in $\chi^2$ would be less than $\sim0.3\%$
probable, given a random sample of data points.

We find that nineteen of our pulsars meet or exceed this sinusoidal
vs. constant-PPA F-Test significance level, indicating statistically
significant sinusoidal PPA variations with time.  It is
important, however, to recognize that a particular pulsar's
F-Test significance calculation is only advisory, given the
uncertainty in the value of the pulse jitter variance (see
\S\ref{sec:jittervar}), which enters crucially as a scale factor in
the reduced $\chi^2$ calculation.  Nevertheless, there is a clear
pattern of statistically significant improvement in many fits across
our pulsar sample when a sinusoidal model is introduced. The details
of the nineteen pulsars' statistically significant fits, along with
two additional ones (see below), are listed in Table
\ref{table:sinefits}.

To highlight our assessment of the quality of the fits, we group the
81 sinusoidal fits into four different classes (see Table \ref{table:sinefits}
for designations of all pulsars in the three highest classes).  
The top two classes
have formally significant ($P\gtrsim0.997$) fits, while the bottom two
do {\it{not.  ``Class
I'' }}pulsars are those whose sinusoidal fits we find particularly convincing. 
The data and fits
for these four pulsars are shown in Fig.~\ref{fig:classIFits}.
{\it{``Class II''}} consists of those pulsars where we find the fit less
convincing, although still formally significant, for two principal reasons.  
(i) If there is a large spread in the PPA values that seems
to not be accounted for by the sinusoid, possibly indicating
non-sinusoidal variation in the PPA (see an example in
Fig.~\ref{fig:classIIFit}); or (ii)  if the {\it{constant}}-PPA fit itself
appears sufficient  $(\chi_{\nu}^2\la 1)$, therefore calling into question the need for a
sinusoid.
{\it{``Class III''}} pulsars are those that we judge might have 
sinusoidal variation, although the fit was not a
formally significant improvement over the constant-PPA fit. 
Both such pulsars' fits are displayed in Fig.~\ref{fig:classIIIFit}.  
{\it{``Class IV''}} encompasses the
remaining 60 pulsars, which have neither formal nor apparent
sinusoidal tendencies.  See
Figure~\ref{fig:classIVFit} for an example.

\subsubsection{Ultralong-Period Variations in Polarized Position Angle}

In our sinusoidal fits ({\it{cf.}} \S\ref{sec:sine}), none of the
trial fits with long-period variations (longer than the total timespan
of observations) is a significant improvement over the constant PPA
fit at the $3\sigma$ level. Therefore, there is no evidence in our
data of variations in PPA, including ramps, on these ultralong
timescales.  See \S\ref{sec:upperlimits} for a discussion of upper
limits on PPA variations from a (constant) mean on shorter timescales.

\subsection{Upper Limits on Temporal Variations  from the Mean  Polarized Position Angle}

\label{sec:upperlimits}

Table \ref{table:big} presents a full summary of our PPA versus time
measurements for the 81 pulsars. The table is organized into three
principal categories.  (i) {\it{Pulsar Details:}} The first seven
entries for a given pulsar list catalog information such as pulsar
name and $DM$ and $RM$; (ii) {\it{Data Catalog:}} The next four
entries detail our observations, listing total timespan, total number
of sessions, and the number rejected due to the PPA being many
$\sigma_{\Delta\psi(t) }$ from the mean (see \S\ref{sec:discont} for
further details on the rejects); and (iii) {\it{Retained Session
Results}}: These results represent important upper limits derived from
the ``sessions retained'' after the editing process, and are discussed
further below.

The standard deviation of the position angle of a given pulsar,
called $\sigma_{\psi}$ in the ``Retained Session Results'' section of
Table \ref{table:big}, is one of the most important quantities
measured in this experiment.  It represents the typical deviation of
any single PPA measurement from the mean position angle.\footnote{The
quantity $\sigma_{\psi}$ is uniquely determined from the set of
measurements over all $L$ retained epochs, $\Delta\psi(t_j); (
j=1,. . .,L)$. This contrasts with the situation for the individual
session error bars, $\sigma_{\Delta\psi(t) }$ of
Eq. \ref{eqn:sigmapsiml}, which depend on the assumptions discussed in
\S\ref{sec:jittervar} regarding pulse jitter that are untestable with
the current data. } (For the pulsars exhibiting possible sinusoidal
variations, the standard deviation from the sinusoid,
$\sigma_{\psi,\rm sine}$, is also presented in Table
\ref{table:sinefits}).  Together, these results represent, to our
knowledge, the only published systematic measurements of time
variations (or their upper limits) of pulsar polarized position
angles.

\subsubsection{Discontinuous Changes in Polarized Position Angle}

\label{sec:discont}
As noted in \S\ref{sec:upperlimits}, there are a handful of measured
$\Delta\psi(t)$ lying many $\sigma_{\Delta\psi(t) }$ away from the
mean value for a given pulsar.  (See the ``Sessions Manually
Rejected'' column of Table \ref{table:big} for a listing of the
pulsars having ``manually'' rejected sessions, and Table
\ref{table:outliers} for details on these sessions' properties.)
There are only six such rejected sessions among all 81 pulsars, and
never more than one per pulsar.  We have examined the polarized
profile in each such case and see no evidence for instrumental
problems.  These puzzling events, if real, indicate the passage of
ionized, magnetized structures across the line of sight on timescales
shorter than the several-month intervals between observing sessions.


\section{Discussion}

\label{sec:discussion}

As noted above, polarized position angle variations could be caused either by phenomena 
at the pulsar (e.g., precession or magnetospheric adjustments in the emission beam geometry); 
or by interestellar Faraday rotation variability.
The observed measurements and upper limits on position angle changes reported in Tables
\ref{table:sinefits} - \ref{table:outliers} can be translated 
(1) directly into  changes or upper limits thereof in projected spin axis or magnetospheric orientation;
or to (2) changes or upper limits thereof in interstellar magnetic fields via Faraday rotation. We now 
discuss each in turn.  

\subsection{Changes in Orientation of the Pulsar or its Magnetosphere}

\subsubsection{Precession or Pulsar Magnetospheric Changes}

Much theoretical work suggests that the amplitude of the precession
would be smaller than a degree, which would place it below the
threshold of measurability for most pulsars in this experiment
({\it{cf.}} Table \ref{table:big}), and furthermore that it would
quickly damp, at least in the presence of rigid pinning or even the
slow creep of superfluid vortices [e.g., \citet {s77,swc99,l06}].
Nevertheless, there are several observations of periodic changes in
pulse shape and/or in arrival times of isolated pulsars in the
literature, along with ours, that seem to indicate the presence of
precession. In a few of the other cases, the derived amplitude is on
the order of one to several degrees and the period is in the few-yr
range, which are similar to our results.  The best documented example
is PSR B1828-11, whose arrival times and pulse shape vary periodically
and in a correlated fashion, indicating precession with a period on
the order of 1 yr and $0\fdg3$ amplitude \citep{sls00}, or
$\sim3\degr$ \citep{le01,r03} or possibly more \citep{alw06}.\footnote{\citet{let10}
show that pulseshape changes are correlated with spindown changes in this
pulsar,
suggesting a magnetospheric rather than precession explanation.}  PSR
B1642-03 also exhibits both phenomena \citep{cd85,b91}, which have
been modeled as stemming from precession with an amplitude of $0\fdg5
- 0\fdg8$ over several yr \citep{c93,slu01}.$^4$  \citet{ss94} discovered
pulse shape and timing variations in PSR B2217+47, which they modeled
as induced by precession with a timescale $\gtrsim 20$ yr.  PSR
B1557-50 shows a several-year timing and dispersion measure
periodicity (the latter interpreted as possible frequency-dependent
temporal pulse shape variations) which is modeled as precession with
an amplitude of $\sim0\fdg01$ \citep{cuo03}.  \citet{dm97} observed
quasiperiodic timing and profile variations in PSR B0959-54 which
could be caused by precession with a period $\gtrsim$ 2500 d and
amplitude $\leq 0\fdg15$. Quasiperiodicities on several hundred day
timescales in Crab Nebula pulsar arrival times may also indicate
precession \citep{lps88,sfw03}.  \citet{dm96} observed pulse shape,
PPA, and timing variations in the Vela pulsar which they suggested
could result from free precesssion with a 330-d period; and
\citet{dr07} further elaborated the precession model to explain
variations in X-ray morphology of the Vela Pulsar's Pulsar Wind
Nebula. \citet{zx06} have proposed that the Galactic Center Radio
Transient GCRT J1745-3009 could be a neutron star precessing by
$\gtrsim 15\degr$, although an experiment like the current one would
be insensitive to its 77-min period.  Finally, the X-ray emitting
isolated neutron star RX J0720.4-3125 exhibits a wide variety of
correlated periodic phenomena indicative of precession on a 7-8 yr
timescale \citep{h06}.

Our analysis shows that 19 of our 81 pulsars exhibit PPA variations
that are significantly better fit (at approximately the 3$\sigma$
level) by a sinusoidal PPA function than by a constant PPA.  Of these,
we judge four pulsars to exhibit the most convincing evidence for
sinusoidal variations in PPA.  (These are our ``Class I'' pulsars:
B0523+11, B0611+22, B0656+14, and B2053+21; see
Fig.~\ref{fig:classIFits}). Our results indicate that variations, and
even sinusoidal variations, are occasionally present in the pulsar
population.  As a group, these 19 pulsars display sinusoidal periods
mostly grouped around 185 to 450~days, with a peak at 200 days and
outliers at approximately 790, 1050, and 1250~days.  The amplitude of
the variation ranges from 1$^\circ$ to 8$^\circ$ for most of the
significant fits, with a peak at 2$^\circ$ and an outlier at
12$^\circ$.  Note that these sinusoidal results do not include the six
large PPA excursions, rejected from the fits in Table \ref{table:big}
and detailed in Table \ref{table:outliers}; as these large excursions
do not carry the signature of precession even if they are real.

In \S\ref{sec:orientation}b, we briefly discussed rotating
magnetospheric currents as a possible source of periodic PPA changes.
In the absence of well-developed models, we only note here that such
phenomena might provide an alternate explanation for the observed
periodic PPA variations.

\subsection{Interstellar Magnetic Field Variations}

Table \ref{table:outliers} lists the properties of those few PPA measurements lying far from
the mean.  As noted above, these could indicate the presence of magnetized variations in the ISM,
although they might conceivably result instead from an experimental problem.  Under the first 
assumption (magnetized ISM variations causing the deviations), we calculate and list  
in Table \ref{table:outliers} the quantities  $\Delta RM(t)$ (Eq. \ref{eqn:Deltafaradayrot})
and $\Delta RM(t)/ RM \approx f \Delta B_{||}(t) / \langle B_{||}   \rangle$ (Eq. \ref{eqn:deltaRMfrac}).    
However, the great majority of our pulsars shows no evidence whatsoever of PPA 
variations above the noise.  Under the second assumption (that the few large PPA deviations observed are 
spurious), 
our measured upper limits on temporal variations of position angle among ``Retained Sessions'' 
(the quantity $\sigma_{\psi}$ 
of Table \ref{table:big}) can be interpreted as upper limits on magnetized variations, again
via Eqs. \ref{eqn:Deltafaradayrot} and \ref{eqn:deltaRMfrac} where the quantities $\Delta \psi_{Faraday}$ and 
$\Delta RM$ are replaced by their respective statistical analogs,  $\sigma_{\psi}$ and $\sigma_{RM}$.  
Table \ref{table:big} lists these  derived upper limits on $\sigma_{RM}$ 
and $\sigma_{RM}/ RM \  (\approx f \sigma_{ B_{||} }/ \langle B_{||}   \rangle$).

There are occasional reports in the literature of temporal variations in interstellar magnetic fields, 
some of which may have origins similar to ours.
First, there are  the cases of the Vela and Crab Nebula Pulsars \citep{hc85,r88}.  
The variations of $RM$ (and also  of $DM$) are so large and frequent that they are almost certainly 
associated with passage of inhomogeneous nebular SNR material across the line of sight. This 
interpretation is easy to reconcile  with these pulsars'  relative youth and the observed presence 
of an SNR.  There are other reports  in the literature of $RM$  variations toward ordinary pulsars on 
multiyear timescales,  sometimes accompanied by $DM$ variations as well \citep{vET97}, which 
appear to be caused by the passage of a cloud across the line of sight.  \citet{7} measured 
many pulsar $RM$s and compared their results with earlier work wherever possible.  
They noted numerous cases of $RM$s changing significantly with respect to values 
measured $\sim(5-20)$ yr earlier, while the $DM$ tended to vary far less on similar timescales.

Yet our sample of pulsars   exhibits only rare evidence for $RM$ variations over the four  years 
of observation, and {\it{no}} evidence  for sustained changes or long-term trends.  We can suggest  
several possible resolutions   to the disagreement between the rather common $RM$ changes in the 
literature and our finding of essentially steady $RM$s in our sample:  (1) $RM$ variations tend to be 
small on the four-year timescale of our experiment, but grow on decade-long scales.  This would have 
interesting implications for the  fluctuation spectrum of interstellar plasma. However, recall that we 
have some sensitivity to changes on timescales longer than our dataspan, but no evidence was found for 
variations on scales beyond $\sim 10^3$ d. (2)  A subtle nonorthogonal emission mode competition 
process can lead to spurious apparent $RM$ variations  \citep{rET04}, which might grow with time. (3)  
One  of the previously published multiepoch $RM$ measurements was incorrect.   This latter possibility 
cannot be ruled out, since $RM$ measurements are difficult, and since the pairs of 
observations were frequently performed by different groups on different telescopes.  Further 
measurements are required in order to determine the correct explanation.

\section{Conclusions}

We have studied the temporal behavior of polarized position angle in
81 pulsars over a $\sim4$ yr period.  We demonstrate that the
intrinsic pulse-to-pulse jitter of polarized subpulses frequently
dominates the uncertainty in position angle determination.  We
searched particularly for sinusoidal variations, discontinuous
changes, and linear trends in the data.  Nineteen pulsars have
variations in PPA that were significantly better fit (at the 3$\sigma$
level) by a sinuoidal function than by a constant; such variations
could be caused by precession or other cyclical phenomena at the
pulsar.  These pulsars (and especially the four pulsars in our ``Class
I'') are good candidates for further study.  A handful of
discontinuous changes were observed in the sample, which could be due
to the passage of the line of sight across an interstellar region with
significantly different magnetized properties.  No linear variations
(i.e., ``ramps'') of PPA across the timespan of observations were
detected.  For all 81 pulsars, we also calculate upper limits on
random variations of PPA from the mean, which provide numerical
constraints on precession and magetized interstellar variations in our
sample.

\acknowledgements{DGB, JEE, JJM, and JMW have been supported by grants
 from the National Science Foundation.  JEE would like to acknowledge
 support from NSF grants AST 0507367, PHY-0215581 (to the Center for
 Magnetic Self-Organization in Laboratory and Astrophysical Plasmas),
 PHY 0821899, and AST 0907837 to the University of Wisconsin.
 Arecibo Observatory is operated by Cornell University under
 cooperative agreement with the NSF We thank Matthew Frank and Ryan
 Terrien for computational assistance.}

\appendix
 \begin{center}
   {\bf APPENDIX}
 \end{center}

 \section{Equations for Determination of a Target Pulsar's PPA Offset with Respect to a 
 Calibrator Pulsar}

\begin{figure}
\begin{center}
\includegraphics[scale=0.9,trim=0.5in -0.5in 0 0.5in]{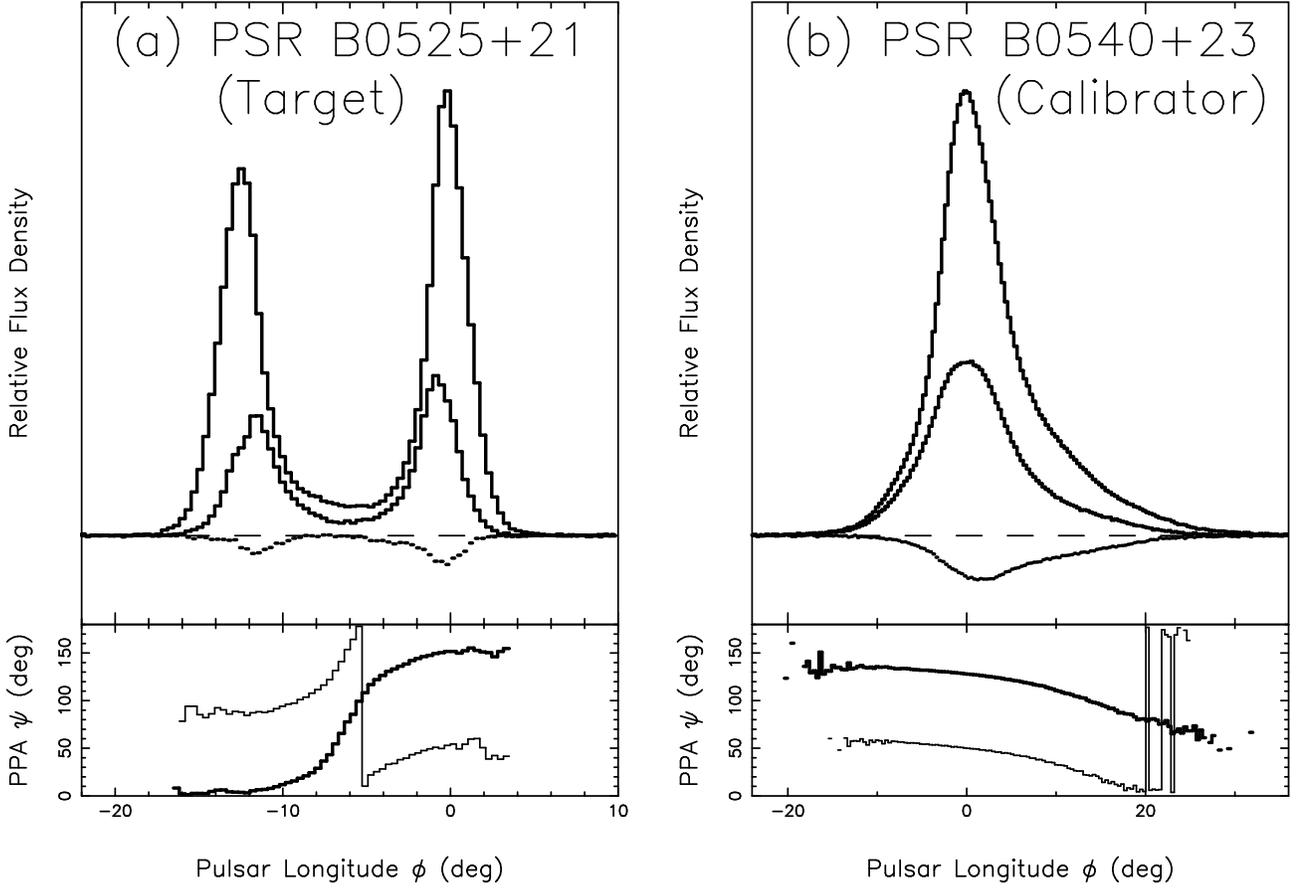}
\caption{
\label{fig:obsPAs}
The polarized position angle (PPA) analysis procedure for target (a) and calibrator (b) pulsars.
In each case, the bottom panel shows measured
PPA as a function of longitude  $\phi$ for a single  session at epoch $t$ (thin line) and for the template 
(thick line), while the top panel shows total flux density (Stokes parameter $I$, top
trace), linearly polarized flux
density ($L$, middle trace), and circularly polarized flux density (Stokes parameter $V$, bottom
trace)   for the template. Note that $180\degr$ position angle jumps result
entirely from the definition of the PPA and have no physical meaning.  {\bf{(a)}} For the target  pulsar, 
we determine the PPA offset between  single session  and template  at each longitude
 ($[\psi (\phi,t)-\bar\psi(\phi)] $ of Eq. \ref{eqn:Deltapsipsr}), and then find the weighted
 mean PPA offset across all longitudes, $\psi(t)$.  {\bf{(b)}}  Similarly, but for the 
 calibrator  pulsar,
 we   determine the PPA offset between single session and template  at each longitude
 ($[\psi_{\rm{cal}}(\phi,t)-\bar\psi_{\rm{cal}}(\phi)] $ of Eq. \ref{eqn:Deltapsical}), and then find the
weighted mean PPA offset across all longitudes $\psi_{\rm{cal}}(t)$. }
\end{center}
\end{figure}

At an observing session at epoch $t$,  we measure the position angle curves 
of target   and calibrator pulsars as a function
of longitude $\phi$, $\psi(\phi,t)$ and $\psi_{\rm cal}(\phi,t)$,  
respectively.  The high $S/N$ position angle curves  of \citet{w99}, which are 
coherent sums over all sessions, 
serve as templates against which to measure  shifts in position angle for a given pulsar 
on a particular observing session.  The  position angle curve 
of these target   and calibrator templates
as a function of longitude $\phi$, is
$\bar\psi(\phi)$  and $\bar\psi_{\rm cal}(\phi)$, respectively.  
(See Fig. \ref{fig:obsPAs} for  examples of session and template  position angle
curves of a target and a calibrator pulsar).  

We first determine $\psi(t)$, the  target pulsar's
weighted\footnote{The weighting factors, $w(\phi)$, account for $S/N$ varying with longitude bin.}, 
longitude-averaged  difference between PPA at an observing session at epoch $t$ 
and PPA of its summed template (see Fig. \ref{fig:obsPAs}a for a target pulsar's session and 
template PPA curves):
\begin{equation}
\psi(t)=
\frac {  \sum_{r=1}^{R}{w(\phi) [\psi(\phi,t)-\bar\psi(\phi)} ]    }
                                                     {  \sum_{r=1}^{R} w(\phi)      },
\label{eqn:Deltapsipsr}
\end{equation}
where the summations are over $R$ longitude bins.

Similarly, we also determine $\psi_{\rm cal}(t)$, the  calibrator pulsar's
weighted$^4$, longitude-averaged 
difference  between PPA at an observing session at epoch $t$ and
PPA of its summed template (see Fig. \ref{fig:obsPAs}b for a calibrator pulsar's session 
and template PPA curves):
\begin{equation}
\psi_{\rm cal}(t)=
\frac {  \sum_{r=1}^{R}{w(\phi) [\psi_{\rm cal}(\phi,t)-\bar\psi_{\rm cal}(\phi)} ]    }
                                                     {  \sum_{r=1}^{R} w(\phi)      }.
\label{eqn:Deltapsical}
\end{equation}

Then the quantity of interest for this experiment, 
$\Delta\psi(t)$, the polarized position angle offset  
of the target pulsar  with respect to a calibrator pulsar  at epoch $t$, is: 
\begin{equation}
\Delta\psi(t)=\psi(t)-\psi_{\rm cal}(t).
\label{eqn:Deltapsiwhole}
\end{equation}

Note then that  the desired $\Delta\psi(t)$  of a target pulsar is ultimately 
determined via  a double-differencing procedure on the various PPAs  -- first both the target
and calibrator pulsars' PPAs at epoch $t$ are differenced from their templates, as in Eqs. \ref{eqn:Deltapsipsr}
and  \ref{eqn:Deltapsical}; and 
finally the  resulting target and  calibrator epoch-template differences are differenced from 
one another,  {\it{cf.}} Eq. \ref{eqn:Deltapsiwhole}.

As pointed out in \S\ref{sec:obsana} (see also Eq. \ref{eqn:basicpsi}), all of the above 
measured position angles 
contain an unknown instrumental origin $\psi_{instr}(t)$.  Note that  
care was taken  to measure both the target  and the  calibrator  pulsar with an  
identical instrumental configuration and hence identical instrumental origin throughout a session at 
some epoch $t$. Therefore $\psi_{instr}(t)$ drops out of Eq. \ref{eqn:Deltapsiwhole} due to the differencing of  two
terms, each containing this quantity. As a result, temporal {\it{variations}} in PPA are accessible even if 
the absolute PPAs themselves are not.

\begin{figure}[h]
\begin{center}
\subfigure
{ \includegraphics[viewport=5 5 500 175,clip=true]{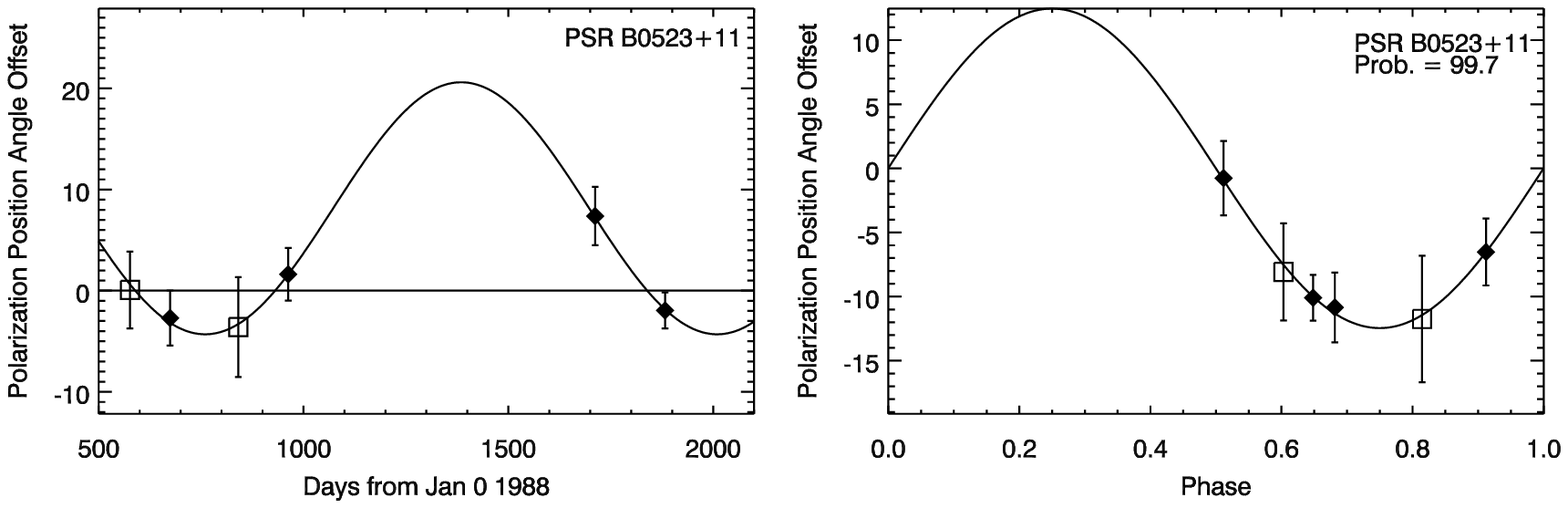} }
\subfigure
{ \includegraphics[viewport=5 5 500 175,clip=true]{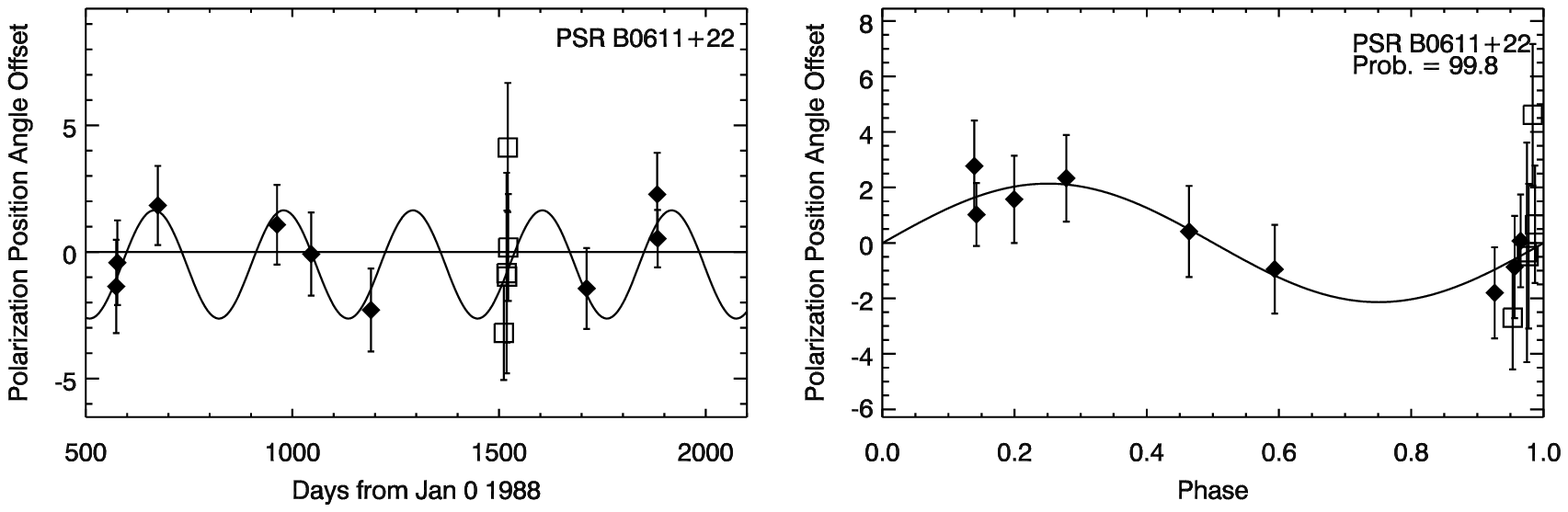} }
\caption{Polarized position angle (PPA) as a function of time for the
four ``Class I'' (best) fits.  For each pulsar, the left panel
presents the data and best sinusoidal fit as a function of epoch,
while the right panel displays the data and fit after folding modulo
the best period.  (The probability that the sinusoid gives a
significant improvement over a constant PPA is listed in the top right
of this panel.)  
The error bars represent the quantity $\sigma_{\Delta\psi(t)}$,
calculated as in Eq.~\ref{eqn:sigmapsiml}.  Normal sessions are
depicted with a filled diamond, while those which required an
artificial composite calibrator are marked with an open square.
\label{fig:classIFits}}
\end{center}
\end{figure}

\begin{figure}[h]
\begin{center}
\includegraphics[viewport=5 5 500
 175,clip=true]{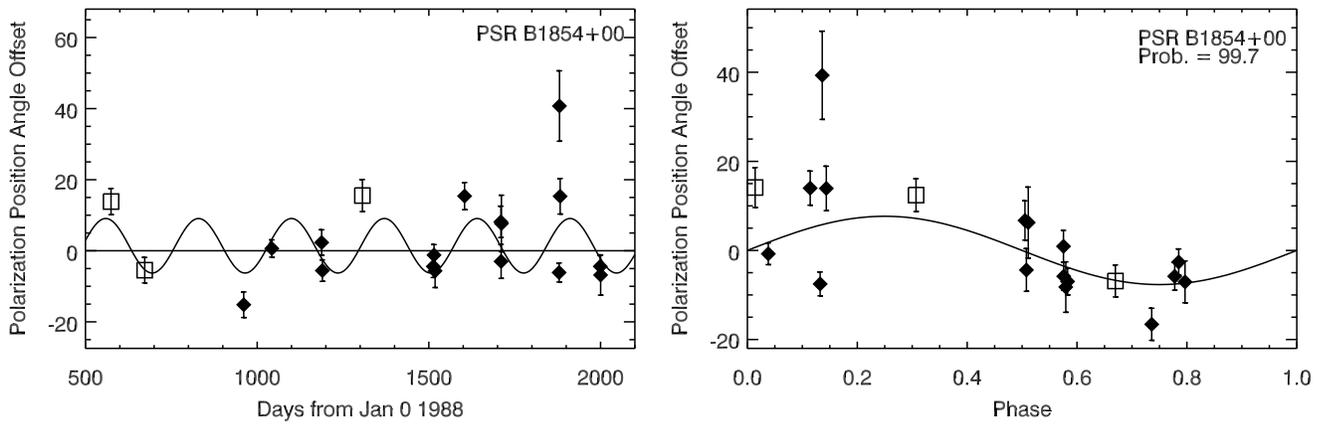}
\caption{As in Figure~\ref{fig:classIFits}, but an example of a sinusoidal fit
that we judged to be ``Class II.'' The fit had similar statistical
significance as Class I fits, but we nevertheless judged it to be less
convincing.  Despite the formal success of the fit, the PPA of
numerous points lies outside the amplitude range of the fitted
sinusoid and another point appears to be entirely out of phase with
it.  See Fig.~\ref{fig:classIFits} caption for further explanations of
the plot labels and symbols.
 \label{fig:classIIFit}}
\end{center}
\end{figure}

\begin{figure}[h]
\begin{center}
\subfigure{
\includegraphics[viewport=5 5 500
 175,clip=true]{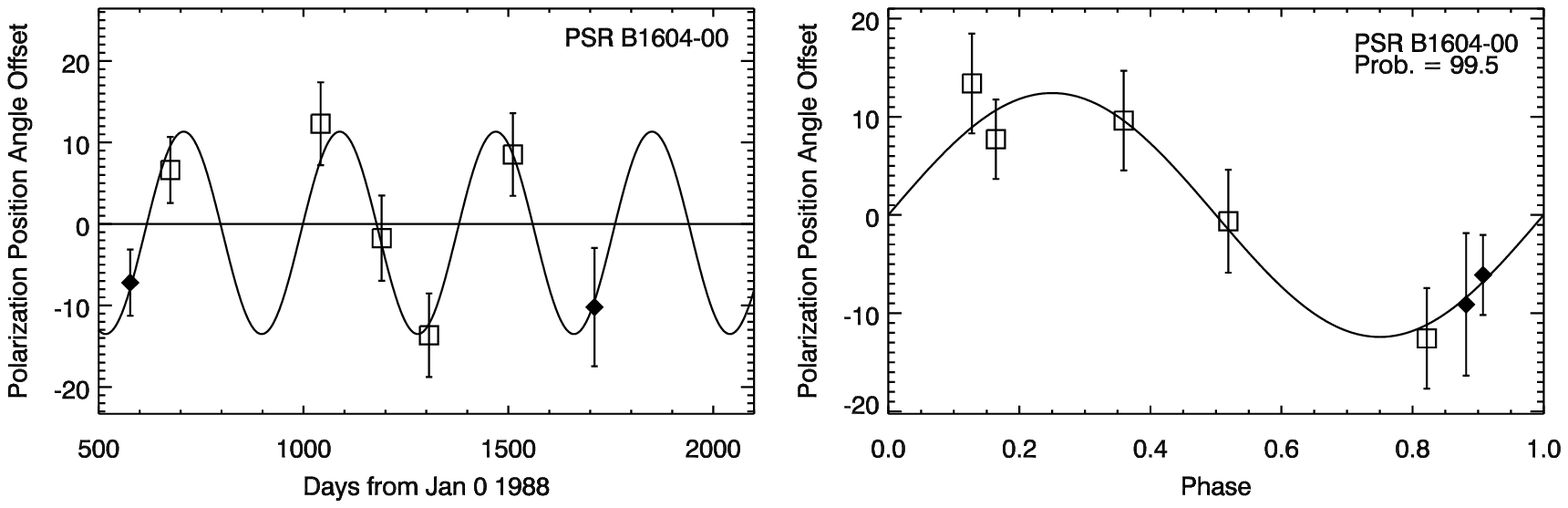}}
\subfigure{
\includegraphics[viewport=5 5 500
 175,clip=true]{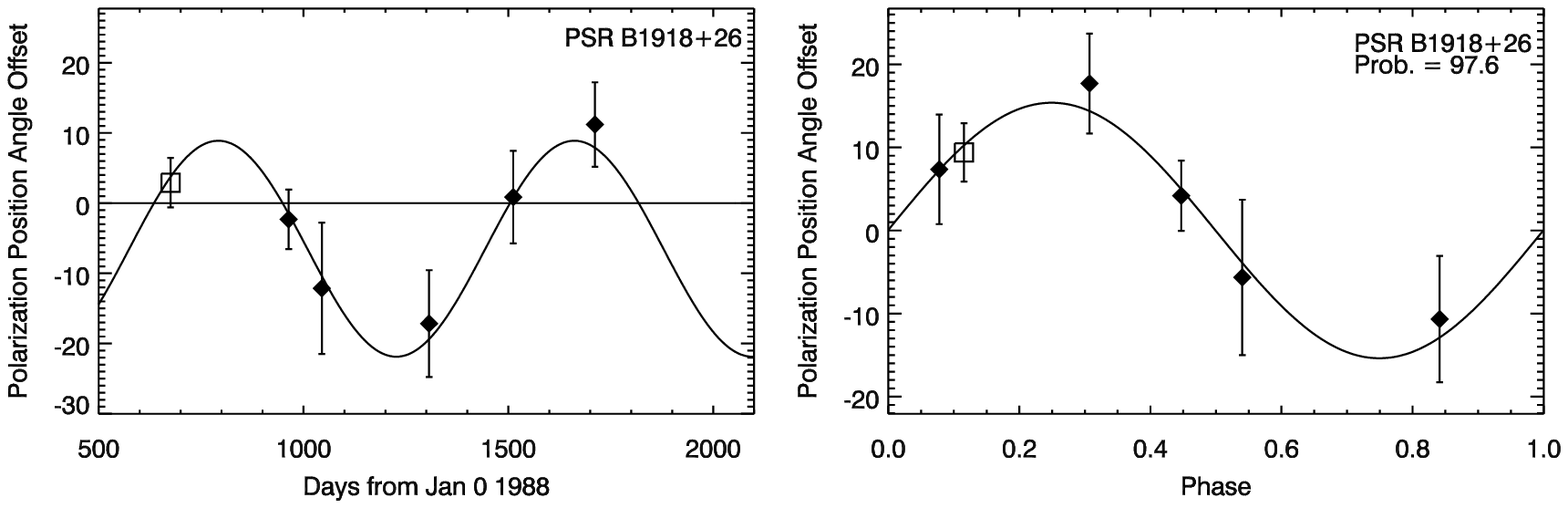}}
\caption{As in Figure~\ref{fig:classIFits}, but for the two pulsars in
 ``Class III.''  Pulsar B1604-00 has a significance-of-fit that is
 near our cutoff (the probability is 0.995).  Given the uncertainties
 in this experiment, and the degree of agreement between the fit and
 observations, we include it as another possible example of
 precession.  This class also include B1918+26; the prefered sinusoid
 fit here is of lower significance (owing partly to one less data
 point than B1604-00), but the correspondence between the fit and data
 points again lead us to flag it as perhaps showing precession.
 \label{fig:classIIIFit}}
\end{center}
\end{figure}

\begin{figure}[h]
\begin{center}
\includegraphics[viewport=5 5 500
 175,clip=true]{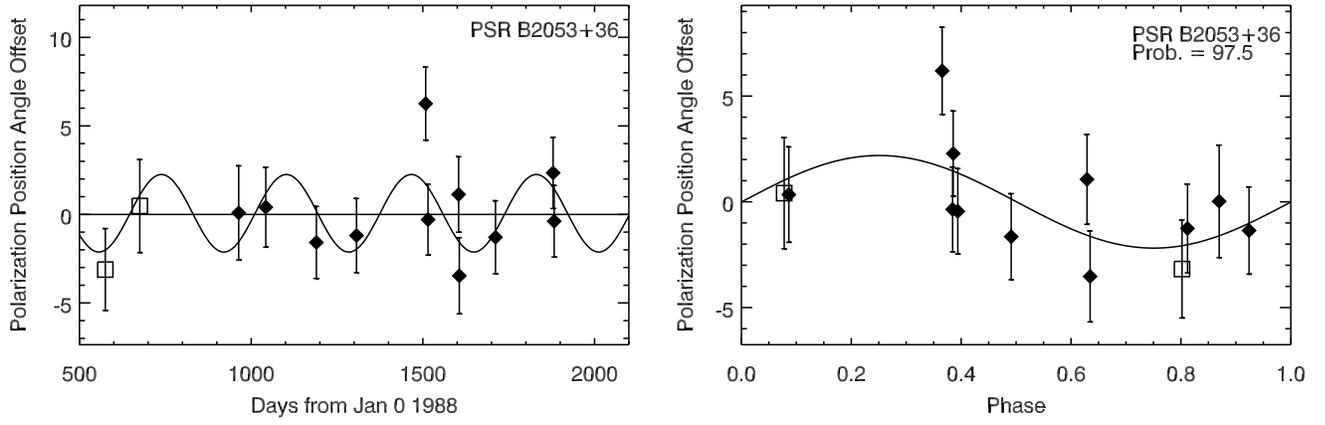}
\caption{As in Figure~\ref{fig:classIFits}, but for a pulsar in
 ``Class IV'': in this case, the sinusoidal fit is not a significant
 improvement over a constant PPA. \label{fig:classIVFit}}
\end{center}
\end{figure}

Further, we focus on temporal {\it{changes}} by 
subtracting a weighted mean offset $\overline{\Delta\psi}$ for each  pulsar,  
before plotting the final values $\Delta\hat\psi(t)$, where

\begin{equation}
\Delta\hat\psi(t)=\Delta\psi(t)-\overline{\Delta\psi}.
\label{eqn:Deltahatpsi}
\end{equation}
and
\begin{equation}
\overline{\Delta\psi}=\frac{     \sum w(t) \Delta\psi(t) }  { \sum w(t)     }, 
\label{eqn:line}
\end{equation}
with the summations being over all observing sessions.  The horizontal
solid line placed at $\Delta\hat\psi=0$ in Figs.
\ref{fig:classIFits} and \ref{fig:classIIFit} then delineates the mean
measured position angle in the new coordinate system defined by
Eq. \ref{eqn:Deltahatpsi}.  The error bar on each individual
measurement is $\sigma_{\Delta\psi (t) }$, as discussed above in
\S\ref{sec:offset} and defined in Eq. \ref{eqn:sigmapsiml}.

\newpage

\begin{deluxetable}{llcrccrcrrrrrr}
\tabletypesize{\scriptsize}
\tablecolumns{14}
\tablecaption{Best Sinusoidal Fits to Polarized Position Angle as a Function of Time }
\tablehead{
\multicolumn{2}{c}{Pulsar} & & Const. & & \multicolumn{9}{c}{Sinusoidal PPA}\\
PSR J & PSR B                    & & PPA \\
\cline{1-2}  \cline{4-4} \cline{6-14} \\
 & & & $\chi^2_{\nu}$ & & $\chi^2_{\nu}$ & $P_{\rm min}$ & $P_{\rm max,i}$ &$P_{\rm best,i}$ & Ampl. & $\sigma_{\psi,\rm sine}$ & F-Test & Prob. & Class\\
& & & & & & [d] & [d] & [d] & $a_{\psi} \ [\degr]$ & $[\degr]$ \\
}
\startdata
J0525+1115 & B0523+11 &     &   1.92 & &     0.03 &  435 & 6532 & 1250 & 12.46 &  0.47 & 312.51 &  0.9968 & I\\
J0614+2229 & B0611+22 &     & 1.01 & &     0.63 &  187 & 6546 &  313 & 2.14 &  1.17 &  10.73 &  0.9982 & I\\
J0629+2415 & B0626+24 &     & 0.63 & &     0.14 &  228 & 5692 &  356 &  2.33 &  0.55 &  35.25 &  0.9997 & II\\
J0659+1414 & B0656+14 &     & 1.38 & &     0.47 &  285 & 7120 & 1047 &  2.47 &  0.85 &  20.24 &  0.9985 & I\\
J1136+1551 & B1133+16 &     & 1.37 & &     0.75 &  234 & 7603 &  364 & 2.81 &  1.47 &  12.97 &  0.9987 & II\\
J1607--0032 & B1604--00 &     & 3.95 & &     0.48 &  324 & 5667 & 380 &12.42 &  3.14 &  46.31 &  0.9948 & III\\
J1740+1311 & B1737+13 &     & 1.05 & &     0.69 &  106 & 6615 &  185 &  2.97 &  1.80 &  15.94 &  1.0000 & II\\
J1805+0306 & B1802+03 &     & 1.98 & &     1.70 &  114 & 7118 &  209 &  2.46 &  2.73 &   6.95 &  0.9980 & II\\
J1823+0550 & B1821+05 &     & 1.25 & &     0.90 &  143 & 7129 &  206 &  4.62 &  2.92 &  10.40 &  0.9995 & II\\
J1844+1454 & B1842+14 &     & 0.60 & &     0.36 &  158 & 7118 &  221 &  1.67 &  1.22 &  14.60 &  0.9999 & II\\
J1857+0057 & B1854+00 &     & 6.08 & &     4.86 &  150 & 7129 &  270 &  7.67 &  7.21 &   7.54 &  0.9974 & II\\
J1915+1009 & B1913+10 &     & 0.41 & &     0.25 &  190 & 7123 &  443 &  1.69 &  1.06 &  12.62 &  0.9993 & II\\
J1916+0951 & B1914+09 &     & 0.23 & &     0.06 &  265 & 6634 &  454 &  1.92 &  0.59 &  27.47 &  0.9993 & II\\
J1920+2650 & B1918+26 &     & 2.25 & &     0.26 &  345 & 5179 &  869 & 15.38 &  2.68 &  41.51 &  0.9764 & III\\
J1932+2220 & B1930+22 &     & 0.82 & &     0.44 &  168 & 7126 &  787 &  1.92 &  1.23 &  16.91 &  0.9999 & II\\
J1954+2923 & B1952+29 &     & 0.83 & &     0.39 &  158 & 7126 &  213 &  4.00 &  1.84 &  22.24 &  1.0000 & II\\
J2022+2854 & B2020+28 &     & 0.89 & &     0.47 &  160 & 7604 &  176 &  4.21 &  1.57 &  18.84 &  1.0000 & II\\
J2037+1942 & B2034+19 &     & 0.75 & &     0.49 &  186 & 6521 &  208 &  6.92 &  3.69 &   9.72 &  0.9974 & II\\
J2055+2209 & B2053+21 &     & 2.69 & &     1.42 &  201 & 6521 &  444 &  5.75 &  3.29 &  13.78 &  0.9990 & I\\
J2124+1407 & B2122+13 &     & 2.13 & &     1.24 &  186 & 6034 &  289 &  8.95 &  4.66 &  11.50 &  0.9980 & II\\
J2212+2933 & B2210+29 &     & 0.70 & &     0.23 &  201 & 6045 &  310 &  3.65 &  1.29 &  26.25 &  0.9998 & II\\
\enddata 
\label{table:sinefits}
\end{deluxetable}
\newpage

\newpage
\begin{deluxetable}{llcccccccccccccc}
\tabletypesize{\scriptsize}

\setlength{\tabcolsep}{0.02in}
\tablecolumns{16}
\tablewidth{0pc}
\tablecaption{Measured Parameters and Quanities Derived Assuming Constant PPA }

\tablehead{

\multicolumn{7}{c}{Pulsar Details}& \multicolumn{4}{c}{Data Catalog}& \multicolumn{5}{c}{Retained Session Results} \\

\cline{1-7}&\cline{7-10}&\cline{11-15} \\

\multicolumn{2}{c}{Name}  &   \multicolumn{2}{c}{DM}    &   \multicolumn{3}{c}{RM} \\

\cline{1-2}&\cline{2-3}&\cline{4-6} \\

\multicolumn{7}{c}{}    & \colhead{Time-}  &\colhead{Total}&\colhead{Sessions} &\colhead{Ses-}  &     \colhead{$\sigma_{\psi}$}      & \colhead{$\sigma_{RM}$}      & \colhead{$\underline{  \sigma_{RM}  }$} & \colhead{Re-}     & \colhead{Outliers}   \\

\colhead{PSR J}  &   \colhead{PSR B} &  \colhead{Value}  &  \colhead{Ref. } &  \colhead{Value}    & \colhead{$\pm$} & \colhead{Ref.} &  \colhead{span}& \colhead{Ses-} & \colhead{Manually}& \colhead{sions} &\colhead{} &\colhead{} &     \colhead{$|RM|$}   & \colhead{duced}   &\colhead{$>3\sigma$}   \\

\colhead{}& \colhead{}& \colhead{(pc/cm$^3$)}& \colhead{}&\multicolumn{2}{c}{(rad/m$^2$)}&& \colhead{(yr)}& \colhead{sions} & \colhead{Rejected}& \colhead{Retained}&\colhead{(deg)}  & \colhead{(rad/m$^2$)} &     \colhead{}   & \colhead{$\chi^2$}   &\colhead{}   \\
}
\startdata
J0304+1932 &  B0301+19 &  15.74 &  1 &   -8.30 &   0.30 &    6 &   3.6 &  13 &  0 & 13 &     1.49 &   0.59 &    0.071 &     0.6 &  0\\
J0525+1115 &  B0523+11 &  79.34 &  1 &   35.00 &   3.00 &   7 &   3.6 &   6 &  0 &  6 &     3.73 &   1.48 &    0.042 &     1.9 &  0\\
J0528+2200 &  B0525+21 &  50.94 &  1 &  -39.60 &   0.20 &    8 &   3.6 &  16 &  0 & 16 &     3.74 &   1.48 &    0.037 &     0.6 &  0\\
J0614+2229 &  B0611+22 &  96.91 &  1 &   69.00 &   2.00 &    6 &   3.6 &  14 &  0 & 14 &     1.75 &   0.69 &    0.010 &     1.0 &  0\\
J0629+2415 &  B0626+24 &  84.19 &  1 &   69.50 &   0.20 &   7 &   3.6 &  10 &  0 & 10 &     1.51 &   0.60 &    0.009 &     0.6 &  0\\
J0659+1414 &  B0656+14 &  13.98 &  1 &   23.50 &   0.40 &   7 &   3.9 &  10 &  0 & 10 &     1.88 &   0.74 &    0.032 &     1.4 &  0\\
J0754+3231 &  B0751+32 &  39.95 &  1 &   -7.00 &   5.00 &     9 &   3.9 &  11 &  0 & 11 &     4.06 &   1.61 &    0.230 &     1.8 &  1\\
J0826+2637 &  B0823+26 &  19.45 &  1 &    5.90 &   0.30 &    6 &   3.9 &   8 &  0 &  8 &     1.56 &   0.62 &    0.105 &     0.6 &  0\\
J0837+0610 &  B0834+06 &  12.89 &  1 &   23.60 &   0.70 &     9 &   3.6 &   8 &  0 &  8 &     2.52 &   1.00 &    0.042 &     1.0 &  0\\
J0922+0638 &  B0919+06 &  27.27 &  1 &   32.00 &   2.00 &     9 &   3.9 &  10 &  0 & 10 &     2.08 &   0.82 &    0.026 &     1.6 &  0\\
J0943+1631 &  B0940+16 &  20.32 &  1 &   53.00 &  12.00 &     9 &   3.9 &   5 &  0 &  5 &     2.69 &   1.06 &    0.020 &     0.8 &  0\\
J0953+0755 &  B0950+08 &   2.96 &  1 &    1.35 &   0.15 &    10 &   3.9 &  10 &  0 & 10 &     1.71 &   0.68 &    0.501 &     1.2 &  0\\
J1136+1551 &  B1133+16 &   4.86 &  1 &    3.90 &   0.20 &    8 &   4.2 &  13 &  0 & 13 &     2.39 &   0.95 &    0.243 &     1.4 &  1\\
J1239+2453 &  B1237+25 &   9.24 &  1 &   -0.33 &   0.06 &    10 &   3.8 &   9 &  0 &  9 &     1.87 &   0.74 &    2.243 &     0.8 &  0\\
J1532+2745 &  B1530+27 &  14.70 &  1 &    1.00 &   0.30 &   7 &   3.3 &   7 &  0 &  7 &    11.76 &   4.65 &    4.654 &     9.0 &  2\\
J1543+0929 &  B1541+09 &  35.24 &  1 &   21.00 &   2.00 &     9 &   3.1 &   8 &  1 &  7 &     3.41 &   1.35 &    0.064 &     1.3 &  0\\
J1607-0032 &  B1604-00 &  10.68 &  1 &    6.50 &   1.00 &    6 &   3.1 &   7 &  0 &  7 &     9.71 &   3.84 &    0.591 &     3.9 &  0\\
J1614+0737 &  B1612+07 &  21.39 &  1 &   40.00 &   4.00 &     9 &   3.6 &  11 &  0 & 11 &     6.50 &   2.57 &    0.064 &     1.8 &  1\\
J1635+2418 &  B1633+24 &  24.32 &  1 &   31.00 &   4.00 &   7 &   3.6 &  10 &  0 & 10 &     8.74 &   3.46 &    0.112 &     3.4 &  1\\
J1740+1311 &  B1737+13 &  48.67 &  1 &   64.40 &   1.60 &   7 &   3.6 &  26 &  1 & 25 &     2.44 &   0.97 &    0.015 &     1.1 &  0\\
J1805+0306 &  B1802+03 &  80.86 &  1 &  100.00 & 100.00 &        \tablenotemark{a} &   3.9 &  25 &  0 & 25 &     3.21 &   1.27 &    0.013 &     2.0 &  1\\
J1823+0550 &  B1821+05 &  66.78 &  1 &  145.00 &  10.00 &     9 &   3.9 &  20 &  0 & 20 &     3.85 &   1.52 &    0.011 &     1.2 &  0\\
J1825+0004 &  B1822+00 &  56.62 &  1 &   21.00 &  13.00 &   7 &   3.1 &   5 &  0 &  5 &     7.40 &   2.93 &    0.139 &     1.3 &  0\\
J1841+0912 &  B1839+09 &  49.11 &  1 &   53.00 &   5.00 &     9 &   3.9 &  16 &  0 & 16 &     3.89 &   1.54 &    0.029 &     2.4 &  1\\
J1844+1454 &  B1842+14 &  41.51 &  1 &  121.00 &   8.00 &     9 &   3.9 &  18 &  0 & 18 &     1.79 &   0.71 &    0.006 &     0.6 &  0\\
J1848-0123 &  B1845-01 & 159.53 &  1 &  580.00 &  30.00 &    10 &   3.9 &  10 &  0 & 10 &     1.28 &   0.51 &    0.001 &     0.2 &  0\\
J1851+0418 &  B1848+04 & 115.54 &  1 &  100.00 & 100.00 &        \tablenotemark{a} &   3.6 &   4 &  0 &  4 &    16.11 &   6.38 &    0.064 &     3.7 &  0\\
J1851+1259 &  B1848+12 &  70.61 &  1 &  158.00 &  16.00 &     11 &   3.9 &  12 &  0 & 12 &     5.27 &   2.09 &    0.013 &     1.4 &  0\\
J1850+1335 &  B1848+13 &  60.15 &  1 &  146.00 &   8.00 &     11 &   3.9 &  17 &  0 & 17 &     1.64 &   0.65 &    0.004 &     0.5 &  0\\
J1856+0113 &  B1853+01 &  96.74 &  1 & -140.00 &  30.00 &   12 &   3.6 &  14 &  0 & 14 &    12.48 &   4.94 &    0.035 &     3.4 &  2\\
J1857+0057 &  B1854+00 &  82.39 &  1 &  104.00 &  19.00 &   7 &   3.9 &  19 &  0 & 19 &     9.08 &   3.59 &    0.035 &     6.1 &  6\\
J1901+0156 &  B1859+01 & 105.39 &  1 & -122.00 &   9.00 &   12 &   3.9 &   8 &  0 &  8 &     8.14 &   3.22 &    0.026 &     1.9 &  1\\
J1901+0331 &  B1859+03 & 402.08 &  1 & -237.40 &   1.50 &     9 &   3.6 &   9 &  1 &  8 &     1.05 &   0.42 &    0.002 &     0.1 &  0\\
J1903+0135 &  B1900+01 & 245.17 &  1 &   72.30 &   1.00 &     9 &   3.6 &  10 &  0 & 10 &     1.94 &   0.77 &    0.011 &     0.4 &  0\\
J1902+0556 &  B1900+05 & 177.49 &  1 & -113.00 &  11.00 &     9 &   3.6 &  10 &  0 & 10 &     3.06 &   1.21 &    0.011 &     0.5 &  0\\
J1902+0615 &  B1900+06 & 502.90 &  1 &  100.00 & 100.00 &        \tablenotemark{a} &   3.6 &   7 &  0 &  7 &     9.77 &   3.87 &    0.039 &     1.5 &  0\\
J1905-0056 &  B1902-01 & 229.13 &  1 &  100.00 & 100.00 &        \tablenotemark{a} &   3.9 &   7 &  0 &  7 &    10.11 &   4.00 &    0.040 &     2.2 &  0\\
J1906+0641 &  B1904+06 & 472.80 &  1 &  100.00 & 100.00 &        \tablenotemark{a} &   3.3 &   9 &  0 &  9 &     1.85 &   0.73 &    0.007 &     0.5 &  0\\
J1909+0254 &  B1907+02 & 171.73 &  1 &  100.00 & 100.00 &        \tablenotemark{a} &   1.3 &   5 &  0 &  5 &     5.79 &   2.29 &    0.023 &     0.5 &  0\\
J1910+0358 &  B1907+03 &  82.93 &  1 & -127.00 &   7.00 &     9 &   3.9 &  17 &  0 & 17 &     6.53 &   2.58 &    0.020 &     1.1 &  0\\
J1909+1102 &  B1907+10 & 149.98 &  1 &  540.00 &  20.00 &     9 &   3.9 &  11 &  0 & 11 &     2.13 &   0.84 &    0.002 &     0.7 &  0\\
J1912+2104 &  B1910+20 &  88.34 &  1 &  148.00 &  10.00 &     9 &   2.5 &   7 &  0 &  7 &     5.39 &   2.13 &    0.014 &     0.5 &  0\\
J1914+1122 &  B1911+11 & 100.00 &  2 &  100.00 & 100.00 &        \tablenotemark{a} &   3.9 &   9 &  0 &  9 &     7.58 &   3.00 &    0.030 &     2.1 &  0\\
J1913+1400 &  B1911+13 & 145.05 &  1 &  435.00 &  30.00 &     9 &   3.9 &  13 &  0 & 13 &     3.25 &   1.29 &    0.003 &     0.8 &  0\\
J1915+1009 &  B1913+10 & 241.69 &  1 &  431.00 &  22.00 &     9 &   3.9 &  15 &  0 & 15 &     1.61 &   0.64 &    0.001 &     0.4 &  0\\
J1916+0951 &  B1914+09 &  60.95 &  1 &  100.00 &   6.00 &     9 &   3.6 &  10 &  0 & 10 &     1.47 &   0.58 &    0.006 &     0.2 &  0\\
J1916+1312 &  B1914+13 & 237.01 &  1 &  280.00 &  15.00 &     9 &   3.9 &  12 &  0 & 12 &     2.97 &   1.18 &    0.004 &     1.1 &  0\\
J1917+1353 &  B1915+13 &  94.54 &  1 &  233.00 &   8.00 &     9 &   3.9 &  13 &  0 & 13 &     2.05 &   0.81 &    0.003 &     0.8 &  0\\
J1918+1444 &  B1916+14 &  27.20 &  1 &  100.00 & 100.00 &        \tablenotemark{a} &   3.6 &  12 &  1 & 11 &     4.13 &   1.63 &    0.016 &     1.6 &  0\\
J1919+0021 &  B1917+00 &  90.31 &  1 &  120.00 &   7.00 &     9 &   2.8 &   5 &  0 &  5 &    12.19 &   4.82 &    0.040 &     2.2 &  0\\
J1920+2650 &  B1918+26 &  27.62 &  1 &  100.00 & 100.00 &        \tablenotemark{a} &   2.8 &   6 &  0 &  6 &     7.91 &   3.13 &    0.031 &     2.2 &  0\\
J1921+1419 &  B1919+14 &  91.64 &  1 &  100.00 & 100.00 &        \tablenotemark{a} &   3.9 &   9 &  0 &  9 &     4.93 &   1.95 &    0.020 &     1.3 &  0\\
J1921+2153 &  B1919+21 &  12.46 &  1 &  -16.50 &   0.50 &     9 &   2.8 &   7 &  0 &  7 &     3.53 &   1.40 &    0.085 &     0.4 &  0\\
J1922+2110 &  B1920+21 & 217.09 &  1 &  282.00 &  14.00 &     9 &   2.5 &   7 &  0 &  7 &    10.13 &   4.01 &    0.014 &     2.2 &  0\\
J1926+0431 &  B1923+04 & 102.24 &  1 &    0.00 &  11.00 &     9 &   2.8 &   8 &  0 &  8 &     2.89 &   1.14 & \tablenotemark{b} &     0.3 &  0\\
J1926+1648 &  B1924+16 & 176.88 &  1 &  320.00 &  14.00 &     9 &   3.6 &  16 &  1 & 15 &     1.33 &   0.53 &    0.002 &     0.2 &  0\\
J1932+2020 &  B1929+20 & 211.15 &  1 &   10.00 &   6.00 &     9 &   2.3 &   4 &  0 &  4 &    10.28 &   4.07 &    0.407 &     1.4 &  0\\
J1933+1304 &  B1930+13 & 177.90 &    3 &  100.00 & 100.00 &        \tablenotemark{a} &   3.6 &   5 &  0 &  5 &    10.09 &   3.99 &    0.040 &     2.4 &  0\\
J1932+2220 &  B1930+22 & 219.20 &  1 &  173.00 &  11.00 &     9 &   3.9 &  17 &  0 & 17 &     1.93 &   0.76 &    0.004 &     0.8 &  0\\
J1935+1616 &  B1933+16 & 158.52 &  1 &   -1.90 &   0.40 &    8 &   3.6 &   8 &  0 &  8 &     0.69 &   0.27 &    0.144 &     0.1 &  0\\
J1937+2544 &  B1935+25 &  53.22 &  1 &  100.00 & 100.00 &        \tablenotemark{a} &   3.9 &  11 &  0 & 11 &     1.25 &   0.49 &    0.005 &     0.2 &  0\\
J1946+1805 &  B1944+17 &  16.22 &  1 &  -28.00 &   0.40 &     9 &   3.6 &   5 &  0 &  5 &     3.52 &   1.39 &    0.050 &     1.9 &  0\\
J1946+2244 &  B1944+22 & 140.00 &    4 &    2.00 &  20.00 &   7 &   2.3 &   4 &  0 &  4 &     7.29 &   2.89 &    1.443 &     0.4 &  0\\
J1948+3540 &  B1946+35 & 129.07 &  1 &  116.00 &   6.00 &     9 &   3.3 &   9 &  0 &  9 &     1.51 &   0.60 &    0.005 &     0.2 &  0\\
J1952+3252 &  B1951+32 &  45.01 &  1 & -182.00 &   8.00 &   7 &   3.6 &  11 &  0 & 11 &     6.55 &   2.59 &    0.014 &     2.8 &  1\\
J1954+2923 &  B1952+29 &   7.93 &  1 &  -18.00 &   3.00 &     9 &   3.9 &  18 &  0 & 18 &     3.04 &   1.20 &    0.067 &     0.8 &  0\\
J2004+3137 &  B2002+31 & 234.82 &  1 &   30.00 &   6.00 &     9 &   3.6 &  15 &  0 & 15 &     2.89 &   1.14 &    0.038 &     0.3 &  0\\
J2018+2839 &  B2016+28 &  14.17 &  1 &  -34.60 &   1.40 &    8 &   3.9 &  20 &  0 & 20 &     3.02 &   1.20 &    0.035 &     0.9 &  0\\
J2022+2854 &  B2020+28 &  24.64 &  1 &  -74.70 &   0.30 &    6 &   4.2 &  19 &  0 & 19 &     2.42 &   0.96 &    0.013 &     0.9 &  0\\
J2030+2228 &  B2028+22 &  71.83 &  1 & -192.00 &  21.00 &   7 &   3.6 &  10 &  0 & 10 &     4.58 &   1.81 &    0.009 &     1.2 &  0\\
J2037+1942 &  B2034+19 &  36.00 &  5 &  -97.00 &  10.00 &   7 &   3.6 &  14 &  0 & 14 &     5.38 &   2.13 &    0.022 &     0.7 &  0\\
J2037+3621 &  B2035+36 &  93.56 &  1 &  100.00 & 100.00 &        \tablenotemark{a} &   3.6 &   7 &  0 &  7 &     4.69 &   1.86 &    0.019 &     1.4 &  0\\
J2046+1540 &  B2044+15 &  39.84 &  1 & -100.00 &   5.00 &   7 &   3.6 &  12 &  0 & 12 &     4.48 &   1.77 &    0.018 &     1.0 &  0\\
J2055+2209 &  B2053+21 &  36.36 &  1 &  -80.50 &   3.00 &   7 &   3.6 &  14 &  1 & 13 &     5.45 &   2.16 &    0.027 &     2.7 &  1\\
J2055+3630 &  B2053+36 &  97.31 &  1 &  -68.00 &   4.00 &     9 &   3.6 &  13 &  0 & 13 &     2.55 &   1.01 &    0.015 &     1.4 &  1\\
J2113+2754 &  B2110+27 &  25.11 &  1 &  -37.00 &   7.00 &   7 &   3.6 &  15 &  0 & 15 &     4.88 &   1.93 &    0.052 &     1.2 &  0\\
J2116+1414 &  B2113+14 &  56.15 &  1 &  -25.00 &   8.00 &     9 &   3.6 &  12 &  0 & 12 &     7.57 &   3.00 &    0.120 &     2.2 &  1\\
J2124+1407 &  B2122+13 &  30.12 &  1 &  -57.00 &   8.00 &   7 &   3.6 &  13 &  0 & 13 &     7.31 &   2.89 &    0.051 &     2.1 &  0\\
J2212+2933 &  B2210+29 &  74.50 &  1 & -168.00 &   5.00 &   7 &   3.3 &  12 &  0 & 12 &     2.80 &   1.11 &    0.007 &     0.7 &  0\\
J2305+3100 &  B2303+30 &  49.54 &  1 &  -75.50 &   4.00 &   7 &   3.3 &  10 &  0 & 10 &     3.57 &   1.41 &    0.019 &     1.3 &  0\\
J2317+2149 &  B2315+21 &  20.91 &  1 &  -37.00 &   3.00 &     9 &   3.3 &   8 &  0 &  8 &     2.79 &   1.10 &    0.030 &     0.7 &  0\\
\enddata 
\tablenotetext{a}{RM unknown; arbitrarily set to 100 for later calculations.}
\tablenotetext{b}{Undefined because RM=0.}
\tablerefs{1: Hobbs et al. (2004b); 2: Hobbs et al. (2004a); 3: Lorimer et al. (2002); 4,   
Hulse \& Taylor (1975); 5, Dewey et al. (1988) ; 6, Manchester (1974); 7, Weisberg et al. (2004);
8, Manchester  (1972);  9, Hamilton \& Lyne (1987); 10, Taylor et al. (1993); 
11,  Rand \& Lyne (1994); 12, Han et al. (2006).   }
\label{table:big}
\end{deluxetable}

\newpage

\begin{deluxetable}{cccrrrcrc}
\tabletypesize{\scriptsize}
\tablecolumns{9}
\tablecaption{Large PPA Offsets and Implied Rotation Measure Offsets}
\tablehead{
\multicolumn{2}{c}{Pulsar} &  \colhead{$t$} & \multicolumn{3}{c}{PPA Offset} & &  \multicolumn{2}{c}{RM Offset}   \\
\cline{1-2}  \cline{4-6}  \cline{8-9}\\
\colhead{PSR J}  &   \colhead{PSR B}          &  \colhead{}&  \colhead{$\Delta \hat{\psi}(t)$}  &  \colhead{$\sigma_{\Delta \hat{\psi}(t)}$} & \colhead{{\underline{$\Delta \hat{\psi}(t)$}}}  & &  \colhead{$\Delta RM(t)$}  &  \colhead{ $\frac{\Delta RM(t)}{RM}   ( \approx  \frac { f \ \Delta B_{||}(t)} {\langle B_{||} \rangle})$ } \\
\colhead{} & \colhead{} & \colhead{(MJD)} &  \colhead{ (deg.)} & \colhead{ (deg.)} &  \colhead{$\sigma_{\Delta \hat{\psi}(t)}$} & &\colhead{(rad/m$^2$)} \\
}

\startdata

    J1543+0929 & B1541+09 & 48870.4 &   -31.4 &     3.0  &  -10.5 & &  -12.4 &   -0.590 \\
    J1740+1311 & B1737+13 & 48678.1 &     7.8 &     1.9  &    4.1 & &    3.1 &    0.048 \\
    J1901+0331 & B1859+03 & 49040.5 &    27.4 &     3.8  &    7.2 & &   10.9 &    0.046 \\
    J1918+1444 & B1916+14 & 48675.1 &   -11.0 &     2.5  &   -4.4 & &   -4.4 &   -0.044\tablenotemark{a} \\
    J1926+1648 & B1924+16 & 49159.4 &   -20.8 &     2.8  &   -7.4 & &   -8.2 &   -0.026 \\
    J2055+2209 & B2053+21 & 48466.8 &    18.5 &     3.2  &    5.8 & &    7.3 &   -0.091 \\
\enddata 
\tablenotetext{a}{RM unknown; arbitrarily set to 100 for this calculation.}
\label{table:outliers}
\end{deluxetable}

\newpage
\newpage
\begin{figure}[h]
\begin{center}
\subfigure
{ \includegraphics[viewport=5 5 500 175,clip=true]{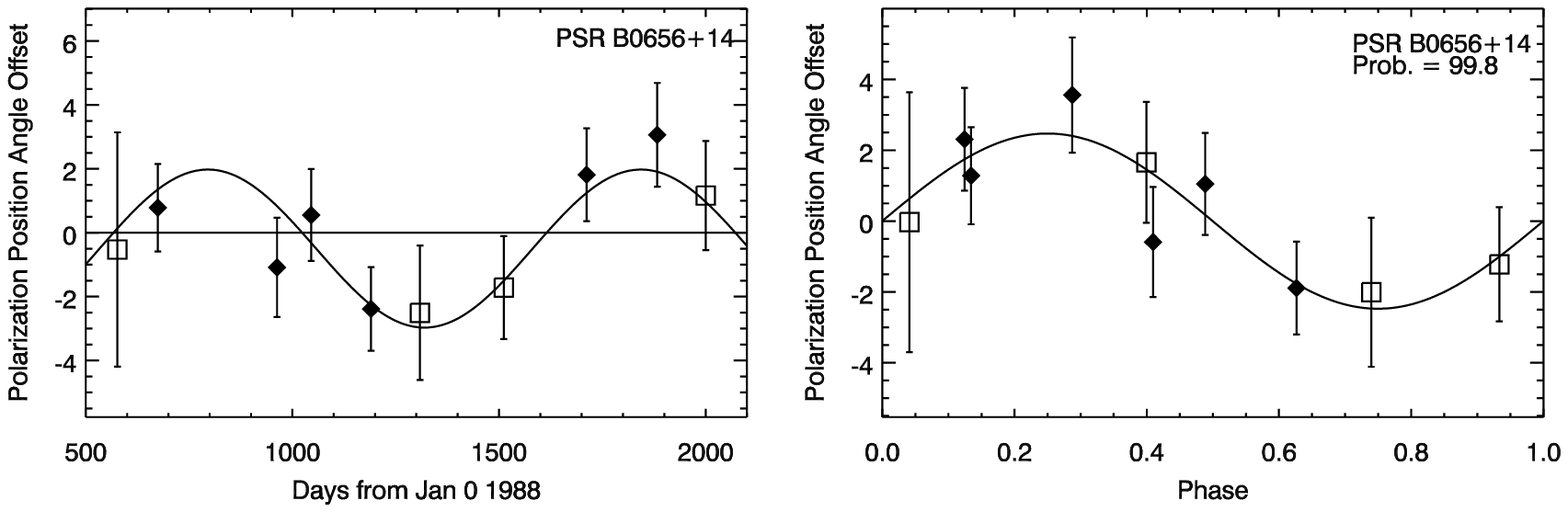} } 
\subfigure
{ \includegraphics[viewport=5 5 500
   175,clip=true]{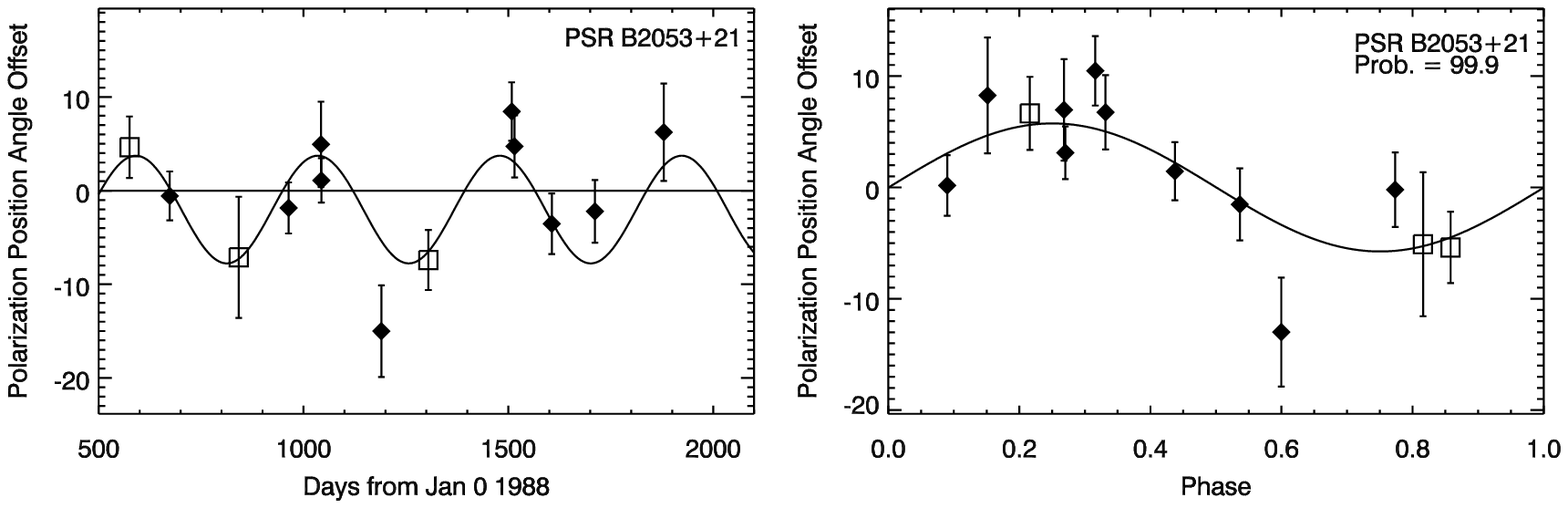} }
\begin{center}
Figure~\ref{fig:classIFits} (continued)
\end{center}
\end{center}
\end{figure}

\end{document}